\pdfoutput=1
\documentclass[11pt,twoside,a4paper,cmspaper,final,collab]{cms-tdr}
\def\svnVersion{aa01ff7}\def\svnDate{2026/11/11}\def\cmsCernNoTag{CERN-EP-2026-118}\def\cmsCernDate{\today}\def\cmsMessage{Published in the Journal of High Energy Physics as \href{http://dx.doi.org/10.1007/JHEP11(2026)063}{\doi{10.1007/JHEP11(2026)063}.}}
\begin{document}\cmsNoteHeader{EXO-24-034}

\newcommand{\PS}{{\HepParticle{S}{}{}}\Xspace}
\providecommand{\Php}{\HepParticle{\PSh}{}{+}\xspace}
\providecommand{\Phm}{\HepParticle{\PSh}{}{-}\xspace}
\newcommand{\Phpm}{\HepParticle{\PSh}{}{\pm}\xspace}
\newcommand{\mS}{\ensuremath{m_{\PS}}\xspace}
\newcommand{\mmumu}{\ensuremath{m_{\PGm\PGm}}\xspace}
\newcommand{\mhh}{\ensuremath{m_{\PSh\PSh}}\xspace}
\newcommand{\mmumuhh}{\ensuremath{m_{\PGm\PGm\PSh\PSh}}\xspace}
\newcommand{\dR}{\ensuremath{\Delta R}\xspace}
\newcommand{\lxy}{\ensuremath{L_{xy}}\xspace}
\newcommand{\sxy}{\ensuremath{\sigma_{xy}}\xspace}
\newcommand{\lxymm}{\ensuremath{L_{xy}^{\PGm\PGm}}\xspace}
\newcommand{\lxyhh}{\ensuremath{L_{xy}^{\PSh\PSh}}\xspace}
\newcommand{\sxymm}{\ensuremath{\sigma_{xy}^{\PGm\PGm}}\xspace}
\newcommand{\sxyhh}{\ensuremath{\sigma_{xy}^{\PSh\PSh}}\xspace}
\newcommand{\mavg}{\ensuremath{\langle m_{\PGm\PGm,\PSh\PSh}\rangle}\xspace}

\providecommand{\cmsLeft}{left\xspace}
\providecommand{\cmsRight}{right\xspace}

\providecommand{\cmsTable}[1]{\resizebox{\textwidth}{!}{#1}}

\cmsNoteHeader{EXO-24-034}

\title{Search for light scalar particles produced in Higgs boson decays in exclusive final states with two muons and two hadrons in proton-proton collisions at \texorpdfstring{$\sqrt{s}=13\TeV$}{sqrt(s) = 13 TeV}}

\date{\today}

\abstract{
	A search for new scalar particles of $\mathcal{O}(\GeVns)$ mass in exclusive final states with muons and light hadrons is presented. The analysis uses proton-proton collision data produced at a center-of-mass energy of 13\TeV collected by the CMS experiment at the CERN LHC in 2016--2018, corresponding to an integrated luminosity of 138\fbinv. The search targets exotic decays of the Higgs boson to a pair of prompt or long-lived identical scalar particles with proper decay lengths $c\tau$ up to 100\mm and masses within the range of 0.4--2.0\GeV. This mass window corresponds to a unique parameter space where hadronic decays of these particles mostly result in a pair of light hadrons. The considered experimental signature is a collimated pair of muons and another pair of charged kaons or pions, each of which may be prompt or displaced. The analysis improves the sensitivity to very light scalar boson masses and demonstrates a novel approach to probe hadronic decays of light scalar bosons. Upper limits on the branching fraction of the Higgs boson to scalar bosons at the level of $\mathcal{O}(10^{-4})$ are obtained for several scalar boson masses between 0.4 and 2.0\GeV, with proper decay lengths of up to $\sim$1\mm.
}

\hypersetup{%
pdfauthor={CMS Collaboration},%
pdftitle={Search for light scalar particles produced in Higgs boson decays in exclusive final states with two muons and two hadrons in proton-proton collisions at sqrt(s) = 13 TeV},%
pdfsubject={CMS},%
pdfkeywords={CMS, EXO, LL}} %

\maketitle 

\section{Introduction}

The discovery of the Higgs boson (\PH) by the ATLAS and CMS experiments~\cite{ATLAS:2012yve,CMS:2012qbp} was a crowning achievement for the standard model (SM) of particle physics, which describes the fundamental particles and their interactions with remarkable precision. The discovery gave rise to the quest to study the properties of this new particle in detail and to confirm or refute its role in the Brout-Englert-Higgs mechanism of electroweak symmetry breaking~\cite{Englert:1964et,Higgs:1964ia,Higgs:1964pj,Guralnik:1964eu} and its Yukawa couplings to fermions.

Many measurements by the ATLAS and CMS Collaborations have since confirmed this theoretical picture with increasing precision~\cite{ATLAS:2022vkf,CMS:2022dwd}. However, only a subset of the couplings of the Higgs boson to SM particles could so far be confirmed, still leaving room for up to $\mathcal{O}(10\%)$ deviations from the SM expectation. Furthermore, the Higgs boson may still decay to invisible particles with branching fractions up to 10.7 and 15.0\%, as constrained by direct searches from the ATLAS~\cite{ATLAS:2023tkt} and CMS~\cite{CMS:2023sdw} Collaborations, respectively. Independent global fits of the SM Higgs boson couplings, with the possibility of the Higgs boson decaying to a potentially visible but undetected final state, have been performed by the ATLAS~\cite{ATLAS:2022vkf} and CMS~\cite{CMS:2022dwd} Collaborations. These fits allow at 95\% confidence level (\CL) the possibility of up to 12 and 16\%, respectively, of the Higgs bosons decaying in this way.

Theoretically, deviations from the SM scalar sector predictions driven by beyond-the-SM (BSM) physics can arise in various ways. A particularly important example consists of the SM singlet $(H^\dagger H)$, where $H$ is the SM Higgs field, coupling to a new real scalar field $S$, itself a singlet, with renormalizable interactions $(\tilde{\mu} S + \lambda S^2)(H^\dagger H)$, where $\tilde{\mu}$ is a mass parameter and $\lambda$ is a dimensionless coupling. 
The scalar field $S$ may then interact with a so-called dark sector of new BSM particles and interactions that also do not carry SM charges. The mixing between the $H$ and $S$ fields originates from the renormalizable interactions, also referred to as the Higgs portal, through which the presence of the dark sector may affect the measured SM Higgs boson properties. In practice, the dark-sector particles acquire an effective coupling to the Higgs boson, yielding a possibly rich phenomenology, depending on their nature and interactions.

In this paper, we consider a minimal Higgs portal scenario with the real singlet field $S$ giving rise to an extra scalar boson \PS that mixes with the SM Higgs boson, as considered in more detail in Refs.~\cite{Clarke2014,Winkler:2018qyg,Boiarska2019}. In particular, light scalar bosons of $\mathcal{O}(1\GeV)$ appear in several SM extensions including cosmic inflation models~\cite{SHAPOSHNIKOV2006414,Bezrukov2009,Bezrukov2013}, supersymmetry~\cite{FAYET1975104,ELLWANGER20101}, dark matter models~\cite{POSPELOV200853,PhysRevD.94.073009}, and cosmological solutions to the hierarchy problem~\cite{PhysRevLett.115.221801}. Many searches have already been conducted for a variety of signatures that result from such scenarios, but usually in the case where the scalar boson is heavy~\cite{FERBER2024104105,CMS:2024zqs}. For the search presented here, we explore the region where the \PS particle has mass $m_\PS \lesssim 2\GeV$, such that hadronic decays of the \PS particle do not result in collimated jets of multiple hadrons, but rather lead dominantly to the exclusive $\PS\to\PKp\PKm$ and $\PS\to\PGpp\PGpm$ decays, respectively, above and below the $m_\PS = 2 \, m_\PKpm$ kinematic threshold. These decays can result in a decay length which is measurable in the detector. In order to maintain the generality of the search, we consider the lifetime of \PS as a free parameter of the model.

\begin{figure}[ht!]
    \centering
    \includegraphics[width=0.5\textwidth]{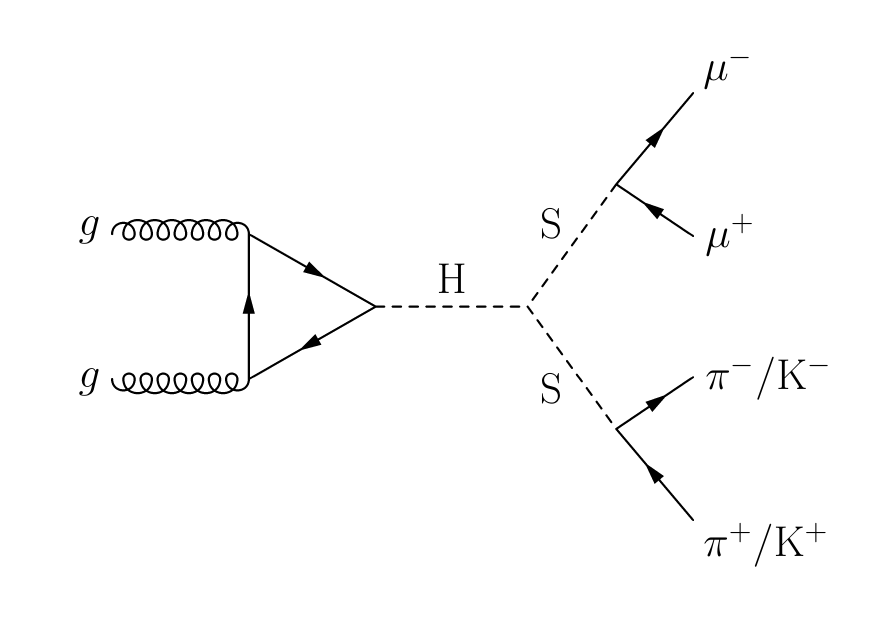}
    \caption{Feynman diagram illustrating Higgs boson-mediated BSM light scalar boson production in the gluon fusion process in the final state of a pair of muons and
    a pair of charged hadrons.}
    \label{fig:feyn_diag}
\end{figure}

A search for low-mass scalar bosons in such exclusive decays was originally proposed in Ref.~\cite{CidVidal:2019urm} for the LHCb Collaboration, using fully hadronic decays. In CMS analyses, without dedicated particle identification methods for high-momentum charged kaons, a fully hadronic final state from inclusive Higgs boson production is overwhelmed by hadronic backgrounds. Instead, we require one of the light scalar bosons to decay to muons, providing a signature that can be triggered efficiently. We consider the $\PH \to \PS\PS \to \PGmp \PGmm \Php \Phm$ signal process, where the hadron \Phpm can be either a charged pion or kaon. A diagram of this process is shown in Fig.~\ref{fig:feyn_diag}.
The values of the $\PS\to\PKp\PKm$ and $\PS\to\PGpp \PGpm$ branching fractions as functions of the light scalar boson mass are taken from Refs.~\cite{Winkler:2018qyg,Gershtein:2020mwi}. The results of the search are interpreted as upper limits on the branching fraction of the Higgs boson into scalar bosons as functions of the scalar boson mass and decay length.

The considered final state has been probed in several prior searches. The search for dimuon resonances at low mass using the CMS scouting data set~\cite{CMS:2021sch} is sensitive to the $\PH \to \PZ_D\PZ_D \to \PGmp \PGmm + X$ decay chain, reaching limits down to $10^{-4}$ on the branching fraction of the Higgs boson to the considered intermediate dark $\PZ_D$ vector boson, depending on the lifetime and the mass of this boson. Of note is that a substantial mass range below $\sim$1.1\GeV was masked out in this search where the presence of SM resonances hindered reliable predictions of the background. A complementary search for a pair of new bosons with masses as low as 0.21\GeV in events with a four-muon final state was also performed by the CMS Collaboration in Ref.~\cite{HIG-21-004}. A search for anomalous showers resulting from decays of new particles in the CMS muon spectrometer also probed new particles with very low masses~\cite{CMS:2024bvl}. The analysis considered a scalar boson \PS decaying to a pair of pions, where the scalar boson originates from a Higgs boson decay. Bounds as low as $10^{-3}$ were achieved on the $\PH \to \PS\PS$ branching fraction for a range of scalar boson masses and lifetimes. Related searches with the considered final state have also been previously conducted by the ATLAS Collaboration~\cite{ATLAS:2022izj,ATLAS:2023cjw}, targeting displaced signatures reconstructed using the calorimeter and/or muon spectrometer for scenarios with dark photons. Several searches for scalar bosons produced in rare \PB-meson decays, targeting $\PS\to\PGmp \PGmm$ decays, have been conducted by the LHCb Collaboration~\cite{LHCb:2015nkv,LHCb:2016awg}, providing constraints on the minimal Higgs portal scenario. 

In the search presented in this paper, we aim for unique, complementary sensitivity to the BSM process depicted in Fig.~\ref{fig:feyn_diag}. The data set consists of proton-proton (pp) collisions at a center-of-mass energy of 13\TeV, collected in 2016--2018 by the CMS experiment at the CERN LHC and corresponding to an integrated luminosity of 138\fbinv. We fully reconstruct the exclusive final state, by identifying the collimated hadron (kaon or pion) and muon pairs from the light-scalar boson decays. By comparing the masses determined from the two track pairs, and requiring that these are isolated, nearly all of the background is suppressed, yielding sensitivity even for prompt signals. 

This paper is structured as follows. In Section~\ref{sec:detector}, the CMS detector is introduced, followed by a description of the simulation samples in Section~\ref{sec:simulation}. The event selection is detailed in Section~\ref{sec:selection}, followed by the estimation of the background in Section~\ref{sec:background}. After a discussion of the systematic uncertainties in Section~\ref{sec:systematics}, we report the results in Section~\ref{sec:results} and summarize in Section~\ref{sec:summary}. Tabulated results are provided in the HEPData record for this analysis~\cite{hepdata}.

\section{The CMS detector} \label{sec:detector}

The CMS apparatus~\cite{CMS:2008xjf,CMS:2023gfb} is a multipurpose, nearly hermetic detector, designed to trigger on~\cite{CMS:2020cmk,CMS:2016ngn,CMS:2024aqx} and identify electrons, muons, photons, and (charged and neutral) hadrons~\cite{CMS:2020uim,CMS:2018rym,CMS:2014pgm}. Its central feature is a superconducting solenoid of 6\unit{m} internal diameter, providing a magnetic field of 3.8\unit{T}. Within the solenoid volume are a silicon pixel and strip tracker, a lead tungstate crystal electromagnetic calorimeter (ECAL), and a brass and scintillator hadron calorimeter (HCAL), each composed of a barrel and two endcap sections. Forward calorimeters extend the pseudorapidity coverage provided by the barrel and endcap detectors. Muons are reconstructed using gas-ionization detectors interleaved with the layers of the steel return yoke outside the solenoid. More detailed descriptions of the CMS detector, together with a definition of the coordinate system used and the relevant kinematic variables, can be found in Refs.~\cite{CMS:2008xjf,CMS:2023gfb}.

The silicon tracker used in 2016 measured charged particles within the pseudorapidity range $\abs{\eta} < 2.5$. For nonisolated particles within $\abs{\eta} < 1.4$ and with transverse momentum $1 < \pt < 10\GeV$, the track resolutions were typically 1.5\% in \pt and 25--90 (45--150)\mum in the transverse (longitudinal) impact parameter~\cite{CMS:2014pgm}. At the start of 2017, a new pixel detector was installed~\cite{Phase1Pixel}; the upgraded tracker measured particles up to $\abs{\eta} = 3.0$ with typical resolutions of 1.5\% in \pt and 20--75\mum in the transverse impact parameter~\cite{DP-2020-049} for nonisolated particles of $1 < \pt < 10\GeV$. According to simulation studies~\cite{DP-2017-015}, similar improvements are expected in the longitudinal direction. 

Events of interest are selected using a two-tiered trigger system. The first level (L1), composed of custom hardware processors, uses information from the calorimeters and muon detectors to select events at a rate of around 100\unit{kHz} within a fixed latency of 4\mus~\cite{CMS:2020cmk}. The second level, known as the high-level trigger (HLT), consists of a farm of processors running a version of the full event reconstruction software optimized for fast processing, and reduces the event rate to a few kHz before data storage~\cite{CMS:2016ngn,CMS:2024aqx}. 

Muons are measured in the pseudorapidity range $\abs{\eta} < 2.4$, with detection planes made using three technologies: drift tubes, cathode strip chambers, and resistive-plate chambers. The single-muon trigger efficiency exceeds 90\% over the full $\eta$ range, and the efficiency to reconstruct and identify muons is greater than 96\%. Matching muons to tracks measured in the silicon tracker results in a relative transverse momentum resolution, for muons with \pt up to 100\GeV, of 1\% in the barrel and 3\% in the endcaps. The \pt resolution in the barrel is better than 7\% for muons with \pt up to 1\TeV~\cite{CMS:2018rym}. 

A particle-flow (PF) algorithm~\cite{CMS:2017yfk} aims to reconstruct and identify each individual particle in an event, with an optimized combination of information from the various elements of the CMS detector. The energy of photons is obtained from the ECAL measurement. The energy of an electron is determined from a combination of the electron momentum at the primary interaction vertex as determined by the tracker, the energy of the corresponding ECAL cluster, and the energy sum of all bremsstrahlung photons spatially compatible with originating from the electron track. The energy of a muon is obtained from the curvature of the corresponding track. The energy of a charged hadron is determined from a combination of the momentum measured in the tracker and the matching ECAL and HCAL energy deposits, corrected for the response function of the calorimeters to hadronic showers. Finally, the energy of neutral hadrons is obtained from the corresponding corrected ECAL and HCAL energies. 

The primary vertex (PV) is taken to be the vertex corresponding to the hardest scattering in the event, evaluated using tracking information alone, as described in Section 9.4.1 of Ref.~\cite{CMS-TDR-15-02}.

\section{Simulated samples} \label{sec:simulation}

The $\PH\to\PS\PS$ signal samples are produced at next-to-leading order (NLO) with the \POWHEG2.0~\cite{Nason_2004,Frixione_2007,Alioli_2010,Bagnaschi:2011tu} Monte Carlo (MC) generator for the dominant Higgs boson production mode of gluon-gluon fusion. The vector boson fusion and the associated Z and W boson production modes are also generated at NLO with the same MC generator to account for their contributions to the total Higgs boson production cross section. With the Higgs boson mass set to 125\GeV, a range of BSM scalar boson massses is considered between 0.4 and 2\GeV. 
The proper decay length of the BSM scalar boson is set to several values between 0.1 and 100\mm ($c\tau_{\text{gen.}}=0.1$, 1, 10, 100\mm), to generate signal decays within the tracker detector volume (up to 600\mm decay length in the laboratory frame). To extend the coverage to $c\tau=0.01\mm$ and achieve finer granularity in this range, nearly 40 additional signal samples with proper decay length $c\tau_{\text{rwt.}}$ are obtained by reweighting the events of the generated signal samples. The reweighting factor is obtained by assuming an exponential distribution to model the decay vertex of the scalar bosons in both the intermediate and generated samples.

\begin{table}[tbh!]
    \topcaption{Branching fractions for the BSM scalar boson decays to a muon pair and to a pion or kaon pair~\cite{Gershtein:2020mwi}, in the targeted range of BSM scalar boson masses \mS. For $\mS\leq 1\GeV$, the hadronic decay mode to kaons is not allowed because of kinematic constraints. The charged kaon decay mode dominates over the charged pion decay mode for $\mS\geq 1.1\GeV$. Subdominant decay modes of the BSM scalar boson have not been included.}
    \label{tab:brfrac}
         \begin{tabular}{cccc@{\hspace*{5pt}}cccc}
        \mS (\GeVns) & $\mathcal{B}(\PS\to\PGmp \PGmm)$ & $\mathcal{B}(\PS\to \PGpp\PGpm)$ && \mS (\GeVns) & $\mathcal{B}(\PS\to\PGmp \PGmm)$ & $\mathcal{B}(\PS\to \PKp\PKm)$  \\ \cline{1-3}\cline{5-7}
        0.4 & 0.144 & 0.571 &&  1.1 & 0.040 & 0.307 \\
        0.6 & 0.095 & 0.603 && 1.2 & 0.071 & 0.401 \\
        0.8 & 0.061 & 0.626 &&  1.4 & 0.085 & 0.413 \\
        0.9 & 0.034 & 0.644 && 1.6 & 0.117 & 0.412 \\
        1.0   & 0.010 & 0.654 && 1.8 & 0.181 & 0.377 \\
           &  &  && 2.0   & 0.239 & 0.350  \\ 
        \end{tabular}
      \end{table}

In the mass range of $0.4\leq\mS\leq2.0\GeV$, the hadronic decay mode consists dominantly of decays to a pair of light hadrons~\cite{Gershtein:2020mwi}. Selecting events online on the basis of the hadronic-only decay mode is challenging as triggers based on jets and/or \pt sum of hadronic activity typically have high thresholds. To overcome this limitation, one BSM scalar boson ($\PS_1$) is fixed to decay to a pair of muons. The other BSM scalar boson ($\PS_2$) is set to decay to a pair of charged pions for $\mS\leq 1.0\GeV$ as the kaon decay mode is not possible because of the limited kinematic phase space. For $\mS\geq 1.1\GeV$, the charged-kaon decay mode is dominant according to the studies in Refs.~\cite{Winkler:2018qyg,Gershtein:2020mwi}. Since the charged-pion mode is suppressed for $\mS\geq 1.1\GeV$, there is little gain in sensitivity that can be achieved by analyzing it, so $\PS_2$ is assumed to decay to charged kaons only. There is also a nonnegligible contribution coming from the scalar boson decays to four-pion final states which has not been included in the samples here. The relevant branching fractions for the muon and charged-hadron decay modes are taken from Ref.~\cite{Gershtein:2020mwi} and listed in Table~\ref{tab:brfrac}. 
These decays are simulated using the \PYTHIA8.240~\cite{Sj_strand_2015} generator. The parton showering, hadronization, and underlying event description are also modeled by \PYTHIA8.240 with the CP2 underlying event tune~\cite{2020}. The NNPDF3.1 leading order~\cite{Ball_2017} sets of parton distribution functions are used in the sample generation. The detector response is simulated with a model of the CMS detector using the \GEANTfour~\cite{Agostinelli:2002hh} toolkit. Additional pp collisions in the same or adjacent bunch crossings (pileup) are described by overlaying simulated minimum bias events on the hard interaction with the multiplicity distribution weighted to be consistent with that observed in data. 

In addition to the signal samples, SM simulation samples for quantum chromodynamic (QCD) and top-antitop quark pair (\ttbar) processes are generated at leading order precision with~\PYTHIA8.240 and at NLO precision with \POWHEG2.0~\cite{Bagnaschi:2011tu,Alioli_2010,Frixione_2007,Nason_2004} generator, respectively. The generated events are interfaced with~\PYTHIA8.240 for parton showering, hadronization, and underlying event description. This is followed by the simulation of detector response and pileup, as described above. The SM simulation samples are used for analysis development and to understand the background only. The final background estimation is performed using data.

The samples are generated separately for each era of the data taking, which includes early 2016, late 2016, 2017, and 2018. The 2016 data are split into two periods because the early 2016 data taking was affected by saturation effects in the silicon strip tracker, which were fixed for the late 2016 data taking~\cite{CMS-DP-2020-045}. 

\section{Event selection} \label{sec:selection}

The large mass difference between the Higgs boson and the BSM scalar bosons leads to their large Lorentz boosts and highly collimated decays. This typically results in at least one of the muons being produced with high transverse momentum ($\langle \pt \rangle\geq 40\GeV$) and small angular distance \dR between the two muons (defined by $\dR=\sqrt{\smash[b]{(\Delta\eta)^2+(\Delta\phi)^2}}$, where $\phi$ is the azimuthal angle). An isolated single-muon trigger with a \pt threshold of 24 (27)\GeV has been employed to collect the data in the 2016 and 2018 (2017) data sets. For the BSM scalar bosons with proper decay lengths $\leq$1\mm, the trigger provides a signal efficiency of $\sim$80\%. 
In spite of the optimal \pt threshold, the presence of a subleading muon track biases the isolation around the leading muon and leads to a slightly reduced signal efficiency. For larger $c\tau$ values (10--100\mm), the signal efficiency drops to 25--50\% because the single-muon trigger is designed to select prompt muons and reject events with muons produced with large impact parameters. In the 2018 data set, corresponding to an integrated luminosity of 60\fbinv, the efficiency for the displaced signatures is improved by the inclusion of displaced dimuon triggers that require at least two muons reconstructed in the muon detector only~\cite{CMS:2018rym} with $\pt > 23\GeV$ each. These triggers do not impose a PV constraint or isolation requirements on the muon tracks, improving the sensitivity to both prompt and displaced signatures. This leads to an increase in the signal efficiency of $\sim$10\%, limited only by the larger \pt threshold on the subleading muon. For the 2016 and 2017 data sets, dimuon triggers (prompt and displaced) were also considered. However, these triggers show a low signal efficiency because of the isolation requirements on the subleading muon, the mass constraints on the dimuon system, and the high \pt thresholds for the muons, so they are not used in the analysis.

At least two muons reconstructed offline and two charged hadrons \Phpm are required in each event. The muon collection consists of particles reconstructed as such by the PF algorithm that pass the loose muon identification criteria, as described in Ref.~\cite{CMS:2018rym}. Additional requirements on the muons include $\pt>5\GeV$ and $\abs{\eta}<2.4$. To ensure that the selected events are well above the trigger threshold, the leading muon is required to have a $\pt>26$ (29)\GeV, for the 2016 and 2018 (2017) data set. For the events in the 2018 data set that are triggered by the displaced dimuon triggers, at least two muons with $\pt>26\GeV$ are required. The charged hadrons are identified by the PF algorithm, which assigns the charged-pion mass hypothesis by default. For $\mS\geq1.1\GeV$, the four-momenta of the charged hadrons are redefined using the charged-kaon mass hypothesis. The charged hadrons are additionally required to have $\pt>5\GeV$ and $\abs{\eta}<2.4$. 

In order to reconstruct the BSM scalar boson candidates, oppositely charged muon pairs are formed with a distance of $\dR(\PGm\PGm)<0.4$. For each pair, a vertex fit is carried out using a Kalman vertex filtering algorithm~\cite{CMS:2014pgm}. A valid vertex fit is produced when the fitted position lies within the tracker volume and the fit converges within a set maximum number of iterations. Among the pairs with a valid fitted vertex, the highest \pt pair is selected as the pair of signal muons coming from the BSM scalar boson. Additionally, the muons are then refitted to the dimuon decay vertex to improve the \pt resolution and the overall mass resolution of the reconstructed BSM scalar boson. For identification of the charged-hadron decay vertex, a similar procedure is followed with a vertex fit of oppositely charged \PGppm/\PKpm pairs. The highest \pt pair is retained as the signal candidate. The hadronic tracks are then refitted to the dihadron decay vertex. To suppress the presence of vertices arising from low-energy pileup, events with the \pt of the dihadron candidate less than 20\GeV are rejected. The dimuon and dihadron candidates are additionally required to pass a loose mass selection of both \mmumu and \mhh to be less than 5\GeV. The Higgs boson candidate mass, \mmumuhh, is reconstructed as the invariant mass of the four objects ($\mu\mu\Ph\Ph$). Events in the mass range $110<\mmumuhh<140\GeV$ are selected. The signal region (SR) is defined as $122.6<\mmumuhh<127.5\GeV$ while the sidebands of the distribution, $110<\mmumuhh<122.6\GeV$ and $127.5<\mmumuhh<140\GeV$, are used to define the control region (CR) for background prediction. The criteria described so far form the baseline selection. 

Comparing SM simulations and data in the CR, the dominant background is composed of multijet events from QCD processes, followed by a contribution from \ttbar production. The QCD multijet background is dominated by events with heavy-flavor jets, which can yield muon pairs in the jet hadronization. Additional selection criteria are applied to suppress the contribution coming from these background processes and are explained below.

\begin{figure}[t!]
                \centering
                \includegraphics[width=0.49\textwidth]{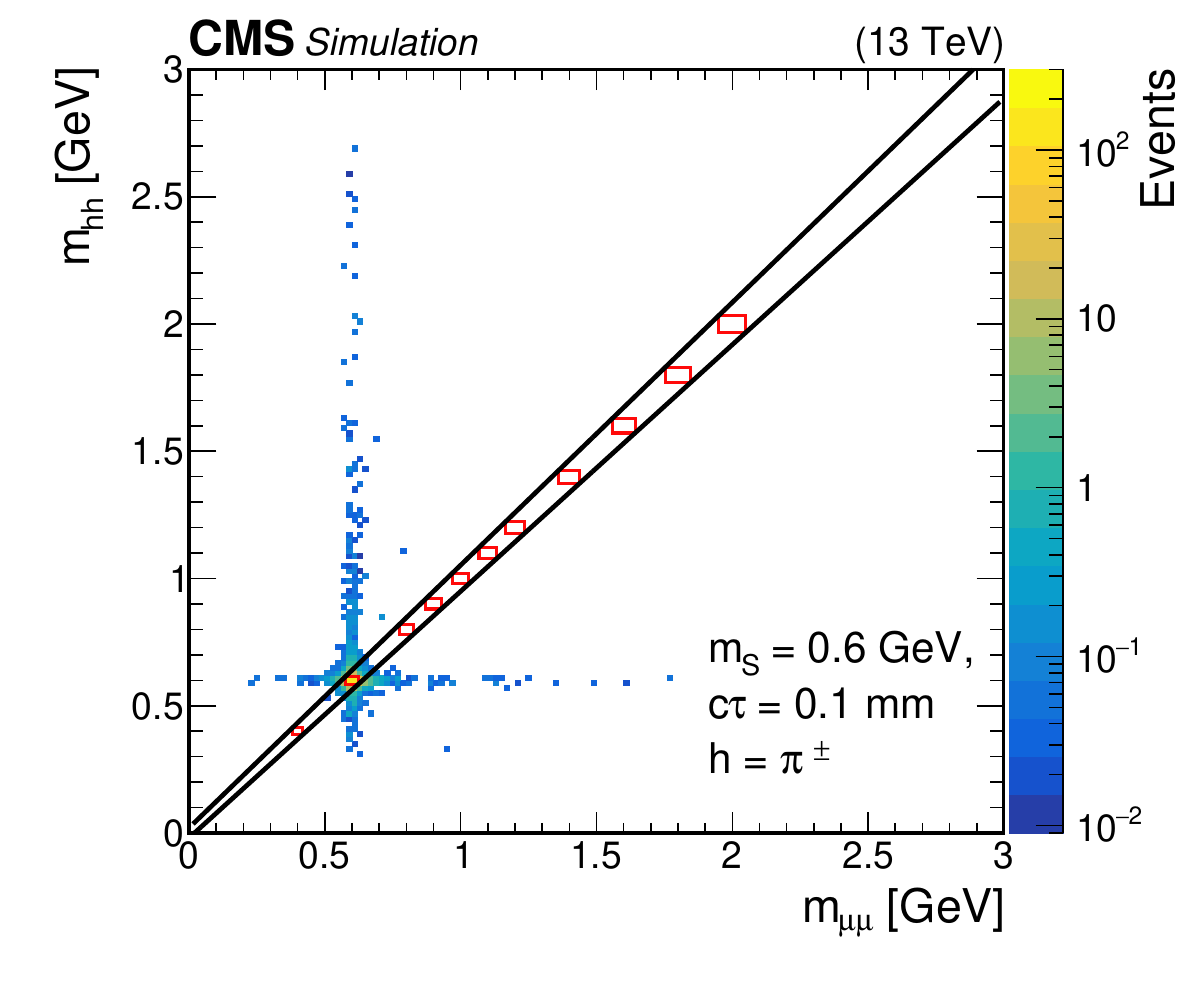}
                 \includegraphics[width=0.49\textwidth]{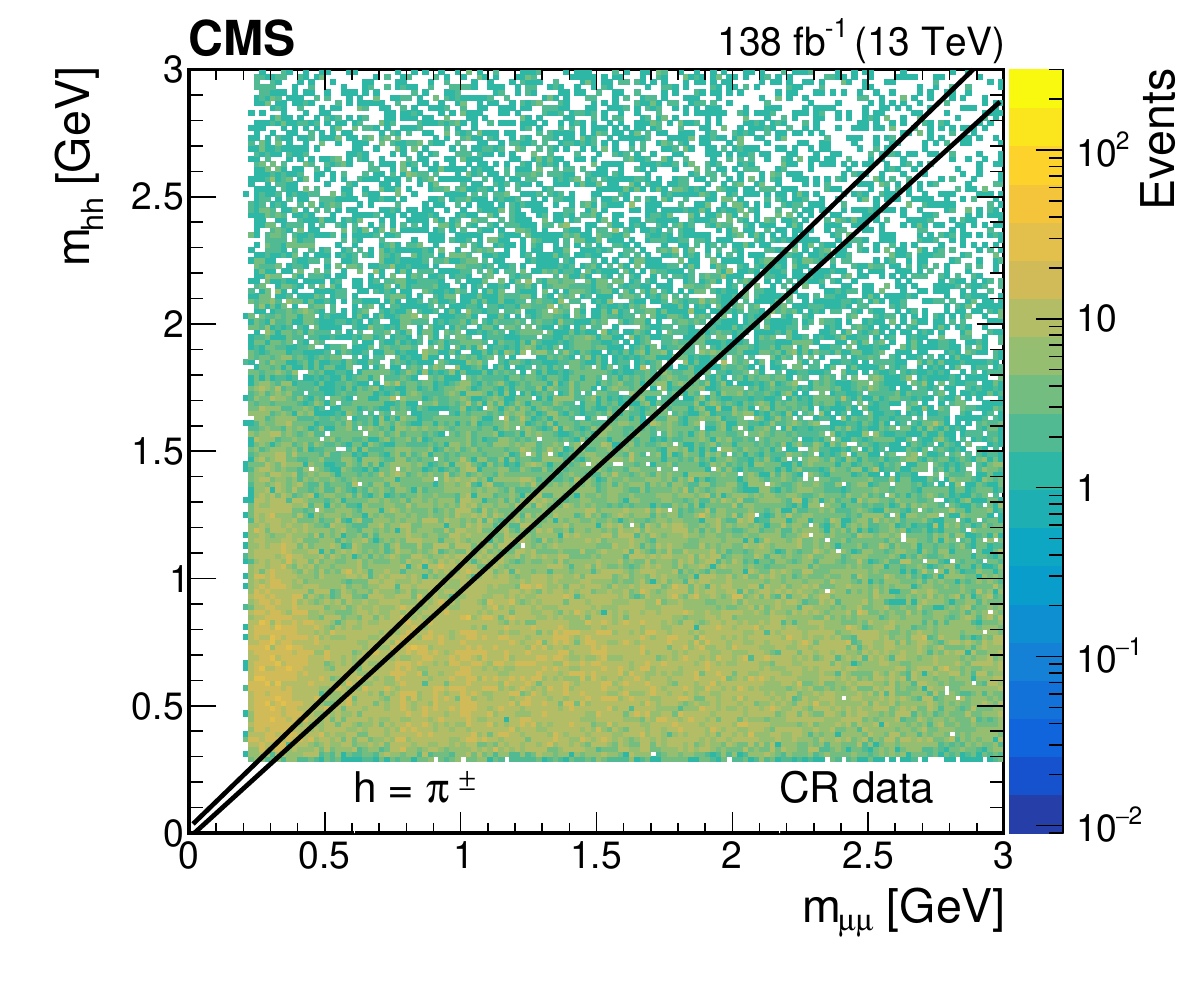}
              \caption{Two-dimensional distribution of the invariant masses \mhh versus \mmumu for a BSM scalar boson of $\mS=0.6\GeV$ and proper decay length $c\tau=0.1\mm$ (\cmsLeft) and the 2016--2018 data set in the CR (\cmsRight), after the baseline selection is applied. The red boxes denote the bounding boxes for each BSM scalar boson mass considered, as discussed in Section \ref{sec:selection}. The black solid lines denote the 2D diagonal mass selection, $\mhh\approx\mmumu$. The distribution of data in the CR (right) is shown for the final state containing charged pions. The boundaries along the dimuon and dihadron mass correspond to the minimum possible dimuon mass ($\approx$0.2\GeV) and minimum possible dipion mass ($\approx$0.3\GeV). For the search targeting charged kaons in the final state, the boundary along the dihadron mass is at the minimum possible dikaon mass ($\approx$1\GeV).}
        \label{fig:masswindow}
\end{figure}

An additional requirement is placed on the dimuon and dihadron invariant masses. The BSM scalar boson is weakly coupled to the SM and is expected to have a small width compared to the detector resolution. Taking into account the detector resolution and final-state radiation for the muons and the hadrons, the signal dimuon and dihadron invariant masses are localized in a narrow region around the BSM scalar boson mass. To exploit this feature, a selection is defined in the two-dimensional (2D) plane of \mmumu and \mhh using all the BSM scalar boson masses considered. This is determined by fitting the dimuon and dihadron invariant mass distributions separately for the prompt-like (i.e., $c\tau=0.1\mm$) signal samples with a Gaussian function on the core of the distribution to reduce the impact of material interactions and reconstruction failures that lead to the non-Gaussian tails. To ensure a high signal efficiency, the fitted mean ($\mu_{\text{fit}}$) and fitted resolution $\sigma_\text{fit}$ of the signal dimuon and dihadron mass distributions are used to construct $\mu_\text{fit}\pm 2.5\sigma_\text{fit}$ bounds separately. In Fig.~\ref{fig:masswindow} (\cmsLeft), the 2D mass plane of $\mmumu$ and $\mhh$ is presented for a BSM scalar boson signal with $\mS=0.6\GeV$ and $c\tau=0.1\mm$, after applying the baseline selection. The extracted bounds for the various signal masses considered are overlaid as red boxes. The edges of the boxes are then linearly interpolated to form the 2D diagonal mass selection that accepts all events falling in it. This selection criterion is applied on all displaced signatures as well. Figure~\ref{fig:masswindow} (\cmsRight) shows the distribution of the expected background estimated from the CR for the 2016--2018 data set, assuming decays to charged pions, where the lines denote the 2D diagonal mass selection. 

A similar number of background events are observed for the search targeting charged kaons in the final state since the redefinition of the charged hadron does not impact the background rejection efficiency. The edge of the dihadron invariant mass distribution however shifts to $\mhh\sim1\GeV$ because of the constraint from the minimum possible dikaon invariant mass. The application of the 2D diagonal mass selection leads to a reduction of the background by $\sim$96\%, therefore providing strong signal vs background discrimination.

\begin{figure}[t!]
 
                \centering
                \includegraphics[width=0.49\textwidth]{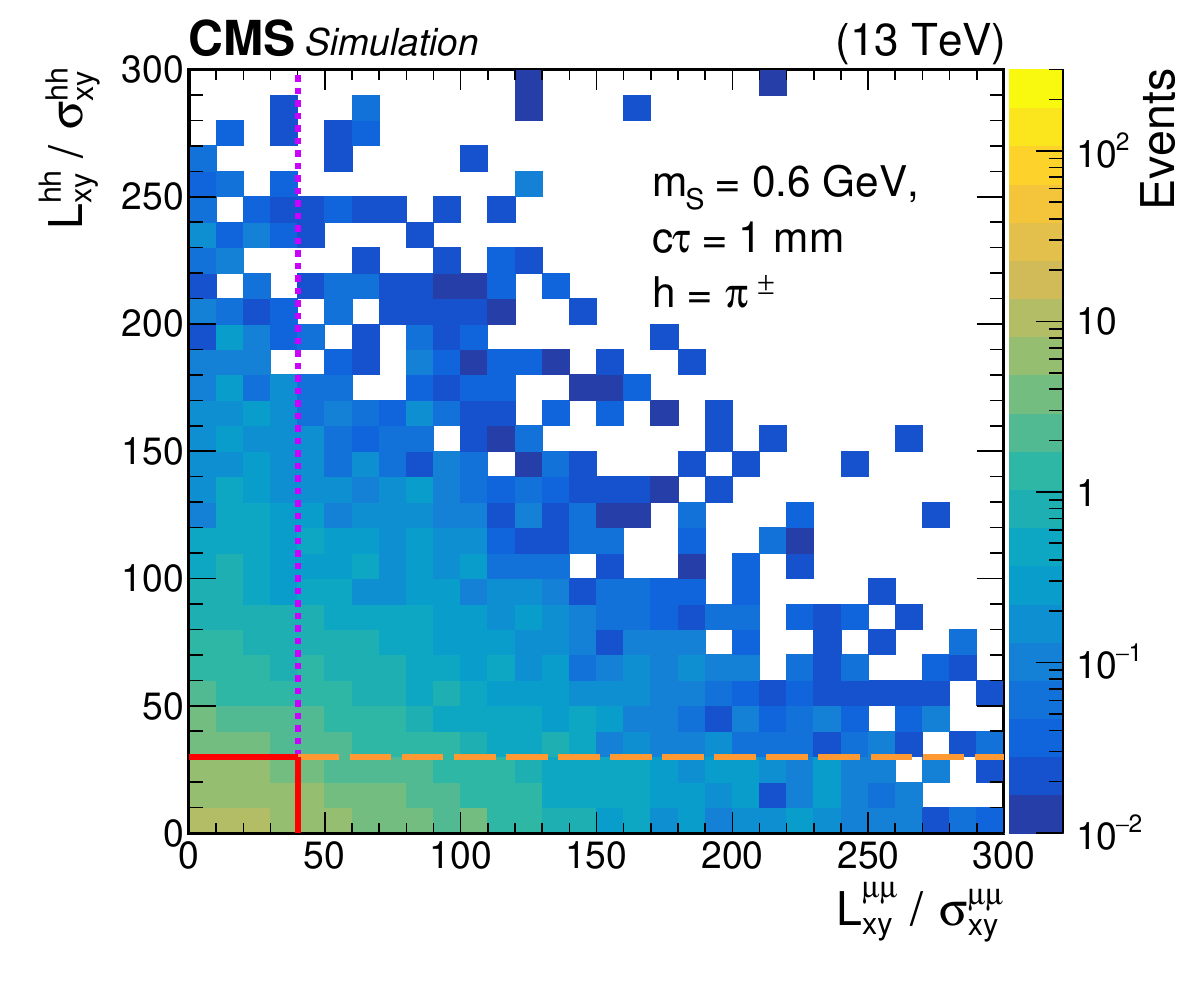}
                 \includegraphics[width=0.49\textwidth]{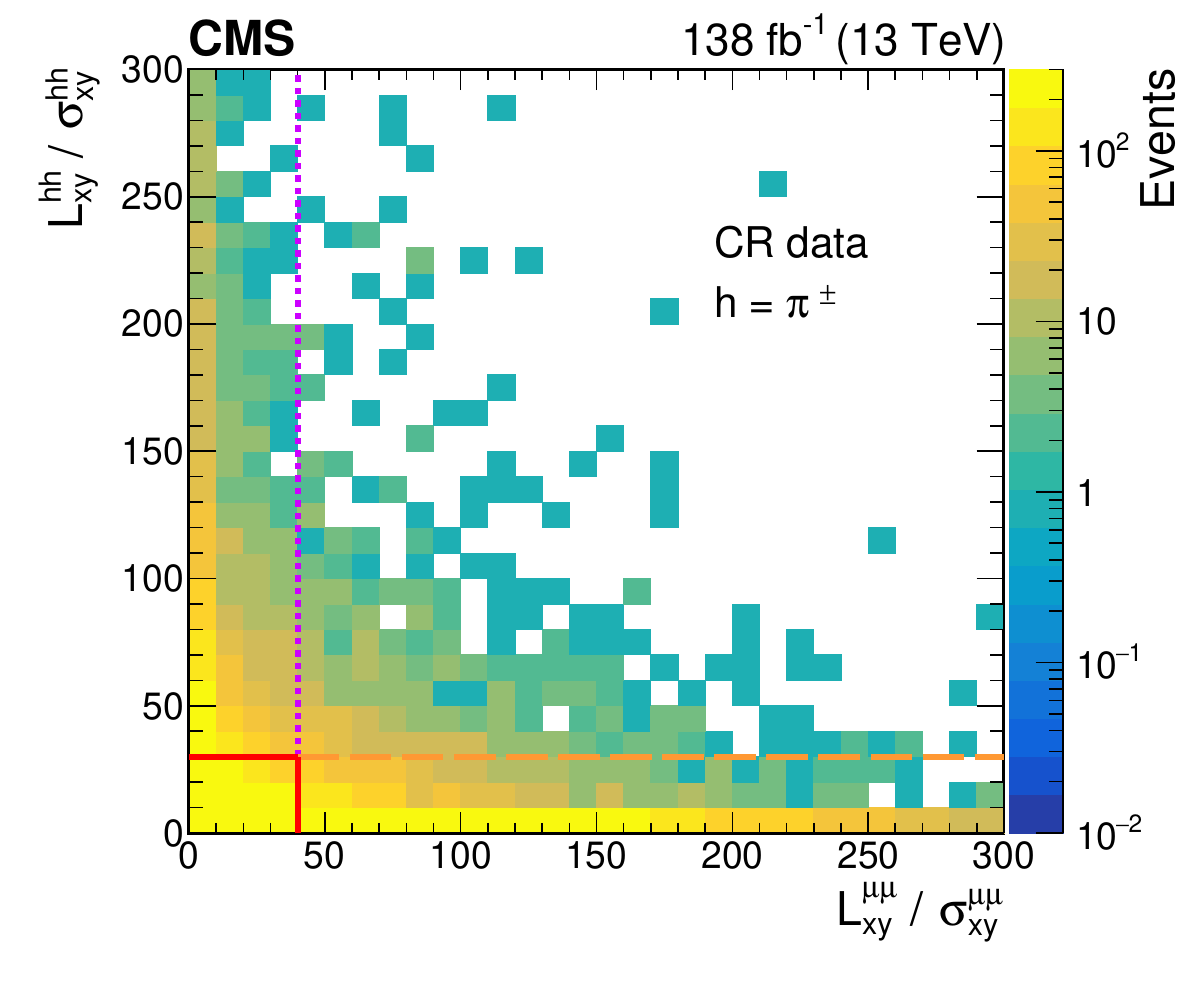}
        \caption{Two-dimensional distribution of the transverse displacement significance $L^{\PSh\PSh}_{xy}/\sigma^{\PSh\PSh}_{xy}$ versus $L^{\PGm\PGm}_{xy}/\sigma^{\PGm\PGm}_{xy}$ for a BSM scalar boson with $\mS=0.6$ \GeV and $c\tau=1\mm$ (\cmsLeft) and the 2016--2018 data set in the CR (\cmsRight) after the baseline selection is applied. The lines denote the boundaries for each of the four categories considered: \textit{prompt} (solid red line), \textit{displaced} $\PGm\PGm$ (dashed orange line), \textit{displaced} $\PSh\PSh$ (dotted pink line), and \textit{displaced} (remaining region). The distribution of data in the CR (\cmsRight) shows that majority of the background is concentrated in the \textit{prompt} category.}
        \label{fig:sig2d_cat}
\end{figure}

The transverse displacement of the decay vertex from the PV (\lxy) and its uncertainty (\sxy) are used to optimize the search for signatures that are either prompt-like or are not prompt-like. This is performed by measuring the transverse displacement significance (defined as $\lxy/\sxy$) for the dimuon and dihadron vertices and splitting the events into categories in the 2D plane of $\lxyhh/\sxyhh$ and $\lxymm/\sxymm$. Figure~\ref{fig:sig2d_cat} shows this 2D plane for $\mS=0.6\GeV$ and $c\tau=1\mm$ (\cmsLeft) and the 2016--2018 data set in the CR (\cmsRight) after the baseline selection is applied. The boundary lines in Fig.~\ref{fig:sig2d_cat} represent the four categories described below: 
\begin{itemize}
        \item \textit{prompt}: $\lxymm/\sxymm<40, \lxyhh/\sxyhh<30$;  This category accepts $>$90\% of the signal for benchmarks with $c\tau< 0.1\mm$. The majority of background lies in this region, since particles in jets arising from QCD multijet processes typically do not exhibit a large displacement. The asymmetric requirements on the dimuon and dihadron transverse displacement significance are chosen to account for the different kinematic constraints of the selected final-state muons and hadrons.
        \item \textit{displaced} $\PGm\PGm$: $\lxymm/\sxymm>40, \lxyhh/\sxyhh<30$; This category captures events where the dimuon vertex is displaced while the dihadron decay is prompt-like.
        \item \textit{displaced} $\PSh\PSh$: $\lxymm/\sxymm<40, \lxyhh/\sxyhh>30$; This category includes events where the dihadron vertex is displaced while the dimuon decay is prompt-like.
        \item \textit{displaced}: $\lxymm/\sxymm>40, \lxyhh/\sxyhh>30$; This category contains events where both dihadron and dimuon vertices are displaced and contributes significantly to the sensitivity for signal benchmarks with $c\tau>1\mm$. The least amount of background is present in this region. 
\end{itemize}

A final selection is placed on the amount of hadronic activity around each selected muon and hadron. This is quantified in terms of the "relative isolation" ($I_{\text{rel}}$) around each object. In particular, the isolation is measured using the \pt contribution coming from reconstructed hadrons (charged and neutral) and photons with $\pt>0.5\GeV$, in the vicinity of the object, i.e., within $\dR < 0.4$. The relative isolation is then defined as the isolation divided by the \pt of the object,
\begin{align}
        I_{\rm{rel}} &= \frac{\Sigma^{\text{ch. had.}}_{\text{not PU}} \pt+\text{max}(\Sigma^{\text{neutr. had.}}\ET+\Sigma^{\gamma} \ET-\Delta \beta,0)}{\pt^{\text{object}}}, 
        \label{eq:pfiso} \\ 
        \Delta\beta &= 0.5\,\Sigma^{\text{ch. had.}}_{\text{PU}}\pt,
\end{align}
where the additional $\Delta\beta$ term is the estimated amount of pileup (PU) contribution in the sum of the transverse energy for the neutral particles in a cone of $\dR=0.4$.

The relative isolation is then calculated for the leading and subleading muons. For measuring the relative isolation of the leading charged hadron, the Eq.~\eqref{eq:pfiso} is modified to exclude the \pt contribution of the subleading charged hadron and vice versa. The relative isolation requirements are determined by maximizing the statistical signal significance, $S/\sqrt{B}$, in the \textit{prompt} category for the BSM scalar boson mass where the largest background is present. A looser set of relative isolation requirements is applied for the non-\textit{prompt} categories (i.e. \textit{displaced} $\PGm\PGm$, \textit{displaced} $\PSh\PSh$ and $displaced$ categories) to recover signal events. The same selections are then applied for all signal samples.

\begin{table}[tbh!]
          \centering
        \topcaption{Summary of the event selection requirements for the analysis. The set of criteria up to (and including) the loose invariant mass row defines the baseline selection. The subscripts 1 and 2 refer to the leading and subleading particles, respectively.}
        \label{tab:selection}
        \cmsTable{
        \begin{tabular}{lccc}
        \rule{0pt}{3.ex}\multirow{2}{*}{Selection} & \multicolumn{2}{c}{\multirow{2}{*}{Requirements}} & Additional \\ 
        & & & information \\ [3pt] \hline 
        \rule{0pt}{3.ex}\multirow{2}{*}{Muons} & \multicolumn{2}{c}{$N_{\mu}\geq 2$, $\pt^{\mu_1}> 26\GeV$,} & $\pt^{\mu_1}>29\GeV$ (2017),  \\ 
        & \multicolumn{2}{c}{$\pt^{\mu}>5\GeV$, $\abs{\eta_\mu}<2.4$} & $\pt^\mu>26\GeV$ (2018 dimuon trigger) \\ [6pt]
        \multirow{2}{*}{Hadrons} & \multicolumn{2}{c}{$N_\Ph\geq 2$, $\pt^\Ph>5\GeV$, $\abs{\eta_\Ph}<2.4$,} & \\
        & \multicolumn{2}{c}{$\Ph=\PGppm$ ($\Ph=\PKpm$ for $\mS\geq 1.1\GeV$)} & \\ [6pt]
        Dimuon & \multicolumn{2}{c}{$\dR_{\PGm\PGm}<0.4$, valid vertex} &\\
        Dihadron & \multicolumn{2}{c}{$\dR_{\Ph\Ph}<0.4$, valid vertex, $\pt^{\Ph\Ph}>20\GeV$} &\\ [6pt]
        Loose  & \multicolumn{2}{c}{$\mmumu<5\GeV$, $\mhh<5\GeV$,} & SR: $\mmumuhh\in[122.6,127.5]\GeV$ \\ 
        invariant mass & \multicolumn{2}{c}{$\mmumuhh\in[110,140]\GeV$} & CR: $\mmumuhh\in[110,122.6)\cup(127.5,140]\GeV$ \\ [6pt] \hline
        \rule{0pt}{3.ex}Di-object & \multicolumn{2}{c}{\multirow{2}{*}{$\mmumu \approx \mhh$}} & 2D diagonal \\
        invariant mass & & & mass requirement \\ [6pt] 
        \multirow{2}{*}{Categories} & \multicolumn{2}{c}{\textit{prompt} ($\lxymm/\sxymm<40, \lxyhh/\sxyhh<30$),} & Non-\textit{prompt} categories made \\
        & \multicolumn{2}{c}{\textit{displaced} $\PGm\PGm$, \textit{displaced} $\PSh\PSh$, \textit{displaced}} & by inverting $\lxy/\sxy$ alternatively \\ [6pt]
         
         & \textit{prompt} & non-\textit{prompt} &  category-wise selections\\ [6pt]
        Relative &  $I_{\text{rel}}^{\mu_1}< 0.3$, $I_{\text{rel}}^{\mu_2}< 0.6$,  & $I_{\text{rel}}^{\mu_1}< 0.5$, $I_{\text{rel}}^{\mu_2}< 0.8$, & $\mu_{1},\mu_{2}=\text{leading, }\text{subleading }\mu$ \\ 
        isolation & $I_{\text{rel}}^{\PSh_1}< 0.6$, $I_{\text{rel}}^{\PSh_2}< 0.8$ & $I_{\text{rel}}^{\PSh_1}< 1$, $I_{\text{rel}}^{\PSh_2}< 1.6$ & $\PSh_{1},\PSh_{2}=\text{leading, }\text{subleading }\PSh$ \\ [3pt] 
                \end{tabular}
        }
       \end{table}

\begin{figure}[th!]
            \centering
            \includegraphics[width=0.49\textwidth]{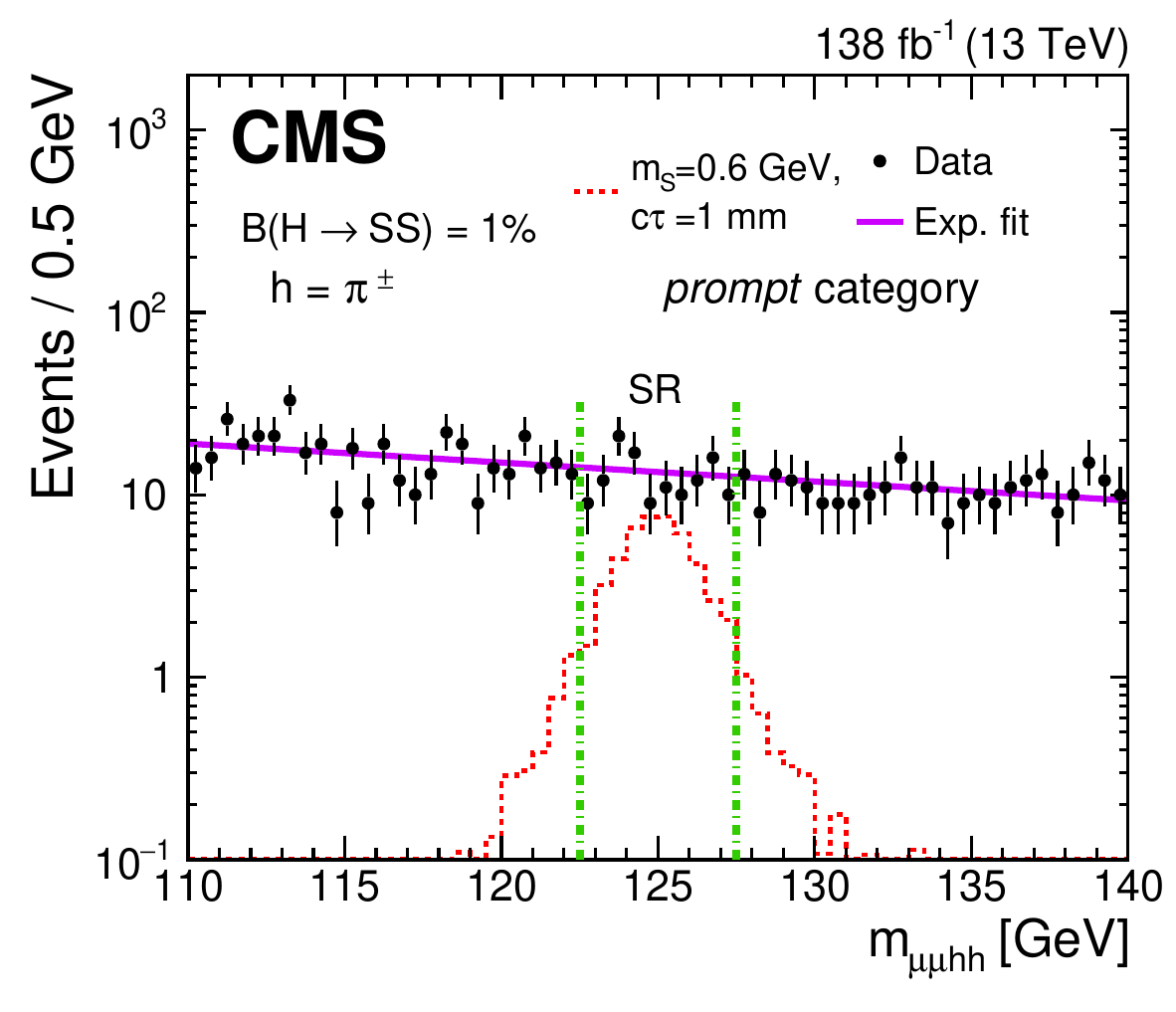}
            \includegraphics[width=0.49\textwidth]{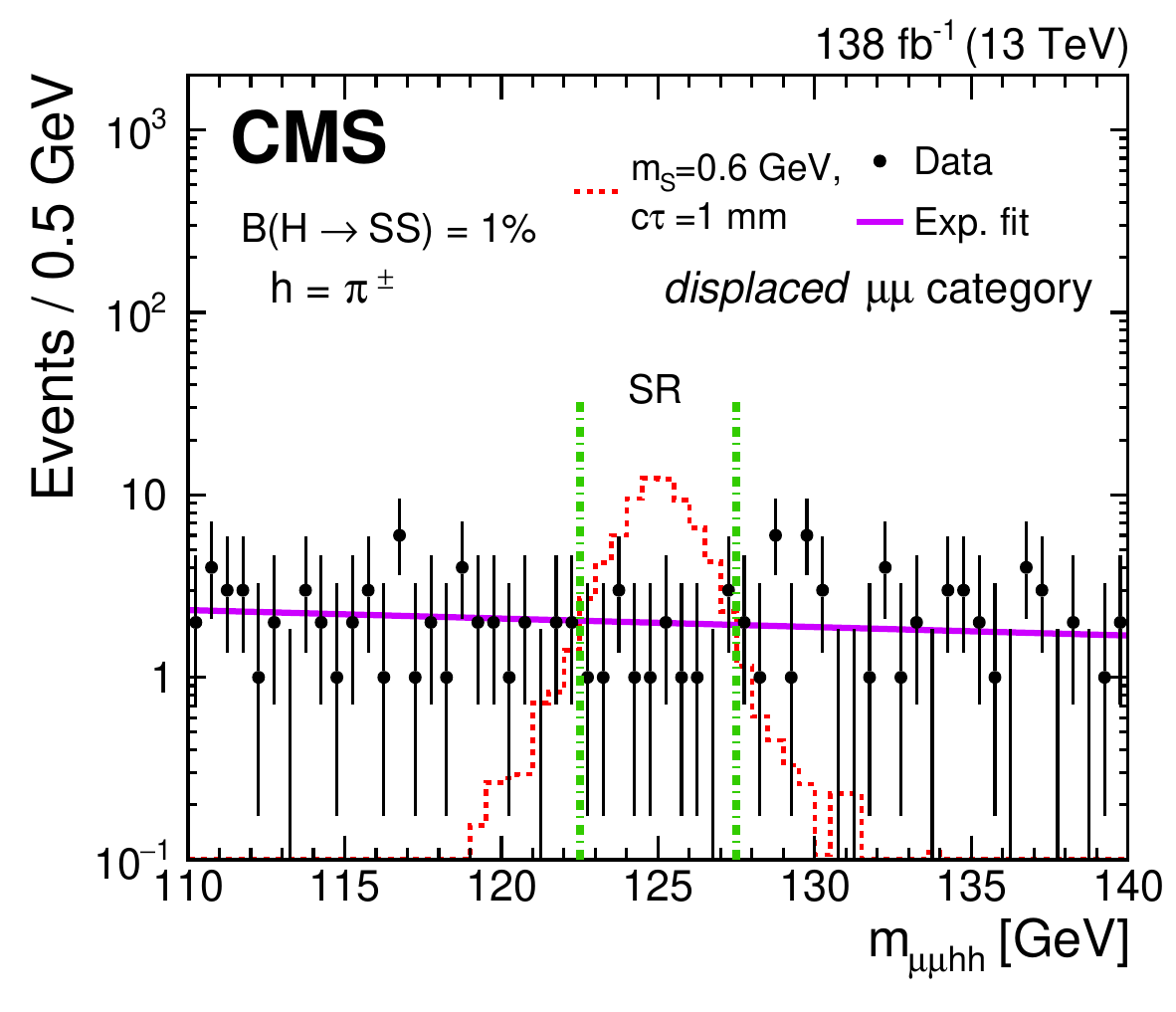}\\
            \includegraphics[width=0.49\textwidth]{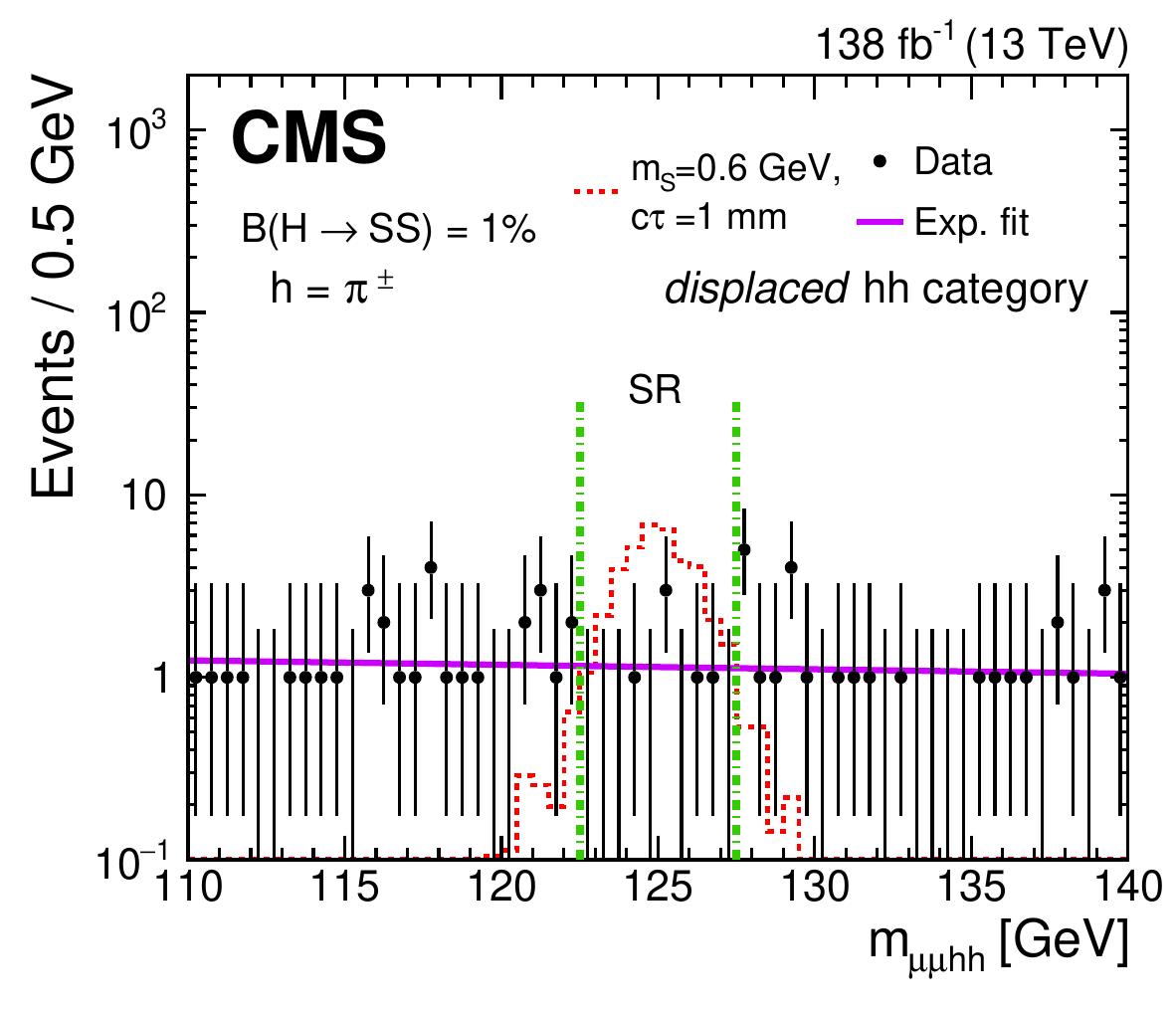}
            \includegraphics[width=0.49\textwidth]{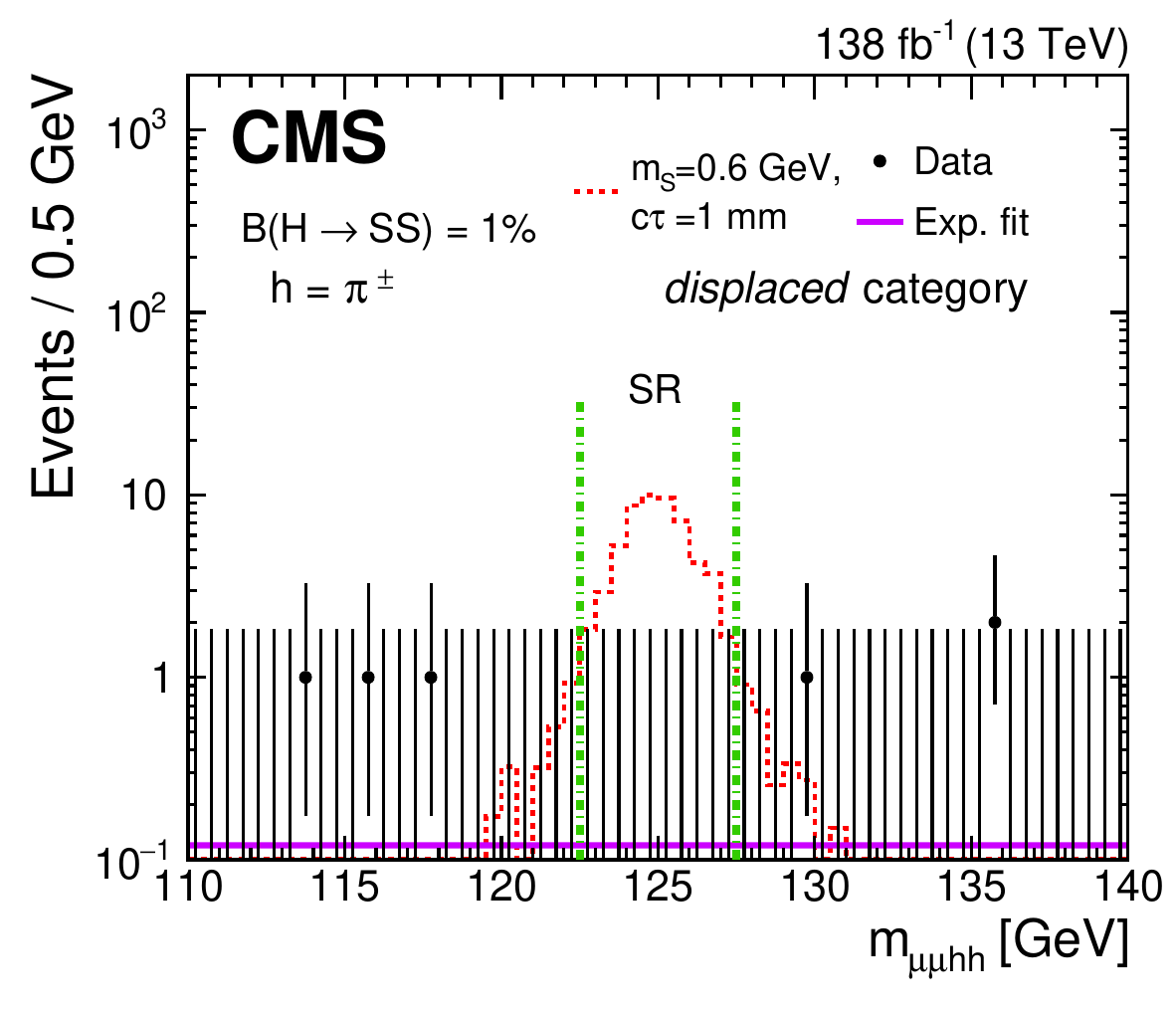}
    \caption{Four-object invariant mass distributions in data (black dots with error bars) and for a BSM scalar boson signal with $\mS=0.6\GeV$ and $c\tau=1\mm$ (red histogram) for the \textit{prompt} (upper left), \textit{displaced} $\mu\mu$ (upper right), \textit{displaced} $\Ph\Ph$ (lower left) and the \textit{displaced} category (lower right). The event selection criteria, as listed in Table~\ref{tab:selection}, have been applied. An additional requirement of $\mavg\in[0.2,2.5]\GeV$ is imposed to display the distribution of data events in the targeted \mS range. The signal event yield is scaled to the integrated luminosity, assuming the gluon fusion Higgs boson production cross section~\cite{ggH_cs}, $\mathcal{B}(\PH\to\PS\PS)=1\%$, and $\mathcal{B}(\PS\to\Pgmp\Pgmm)$, and $\mathcal{B}(\PS\to\PGpp\PGpm)$ for $\mS=0.6\GeV$, as in Table~\ref{tab:brfrac}. The green dash-dotted vertical lines delineate the SR.}
    \label{fig:recohiggsmass_ms0p6_ctau1}
\end{figure}

\begin{figure}[ht!]
\centering
                \includegraphics[width=0.49\textwidth]{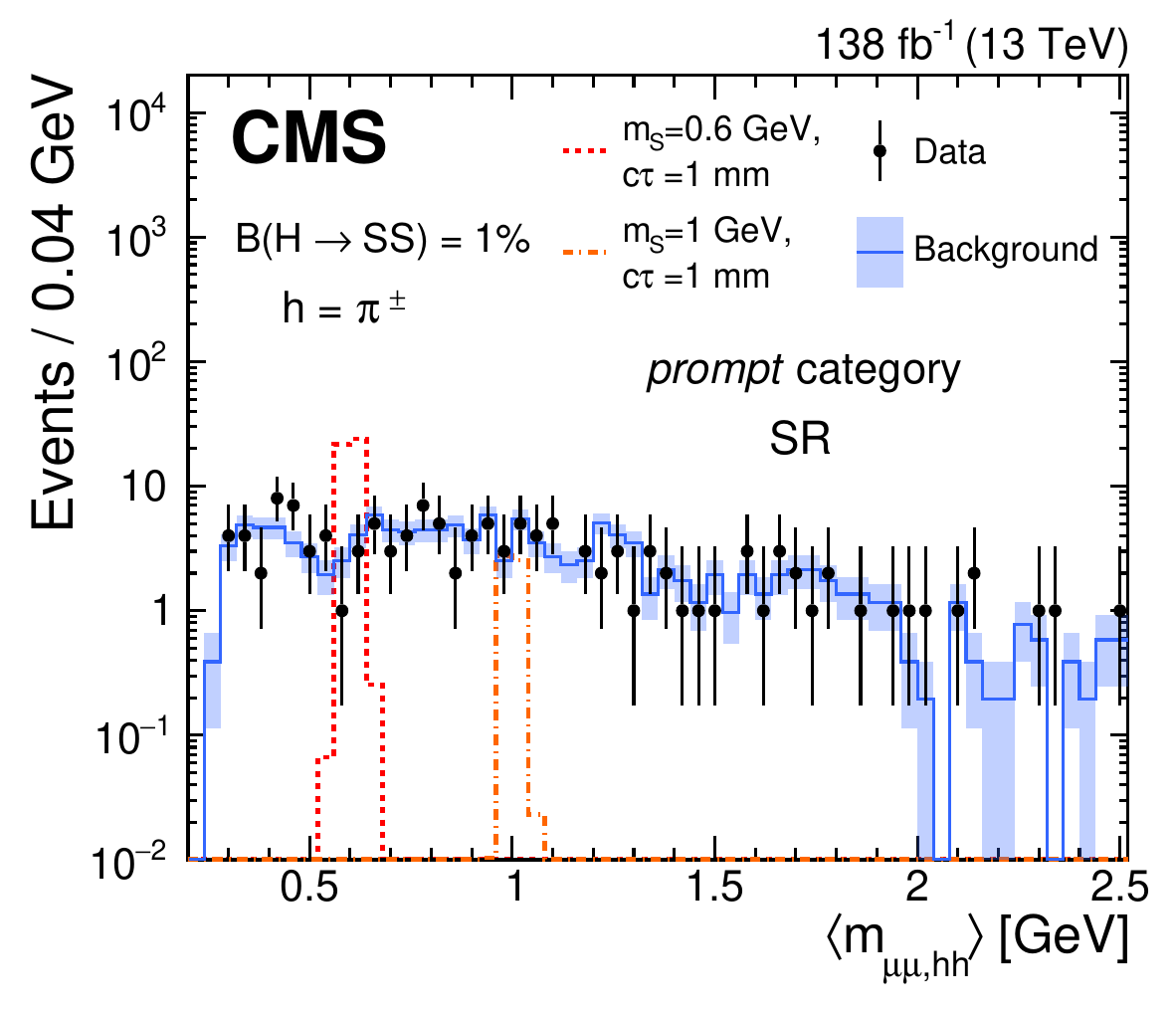}
                \includegraphics[width=0.49\textwidth]{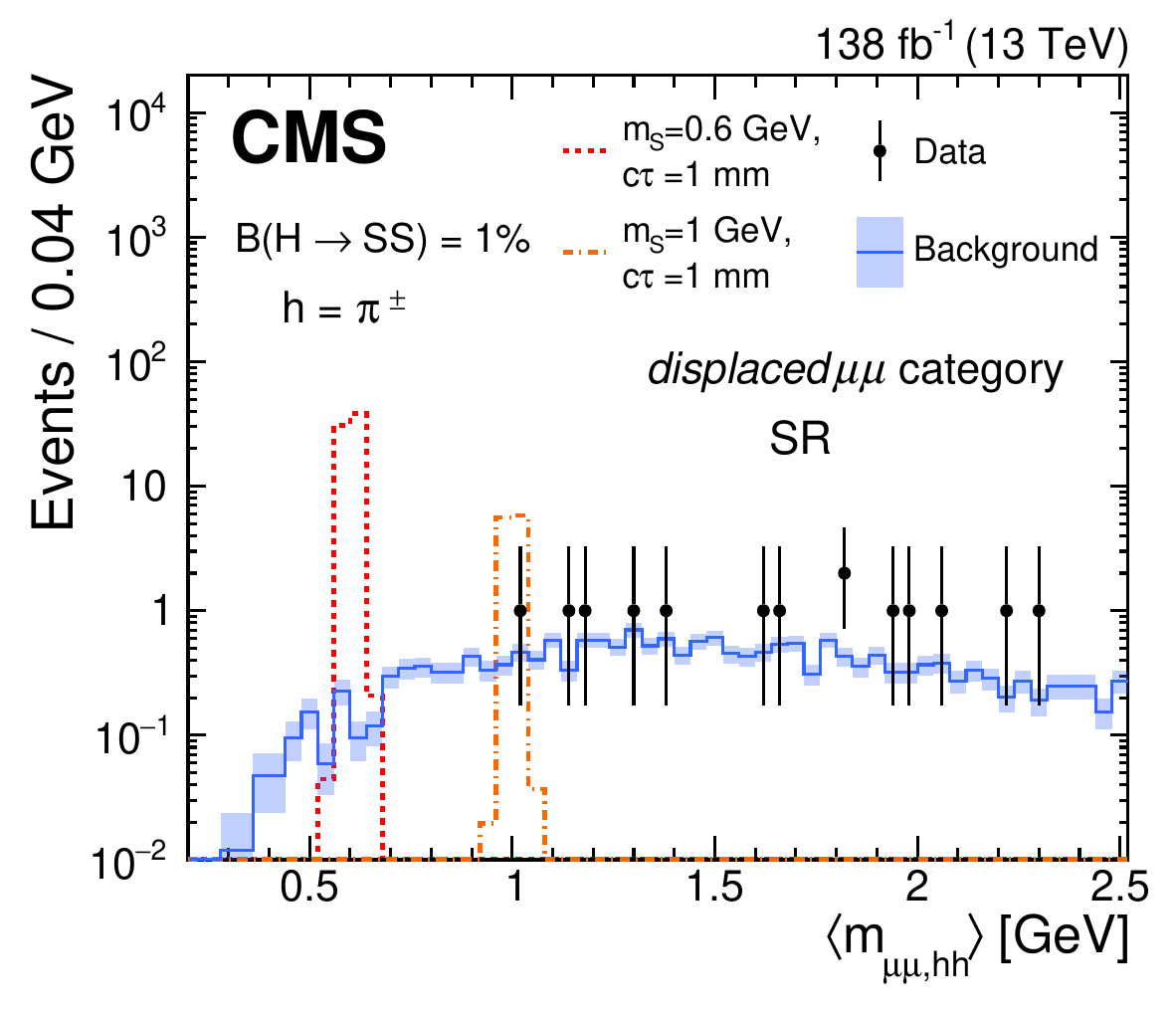}\\
             \includegraphics[width=0.49\textwidth]{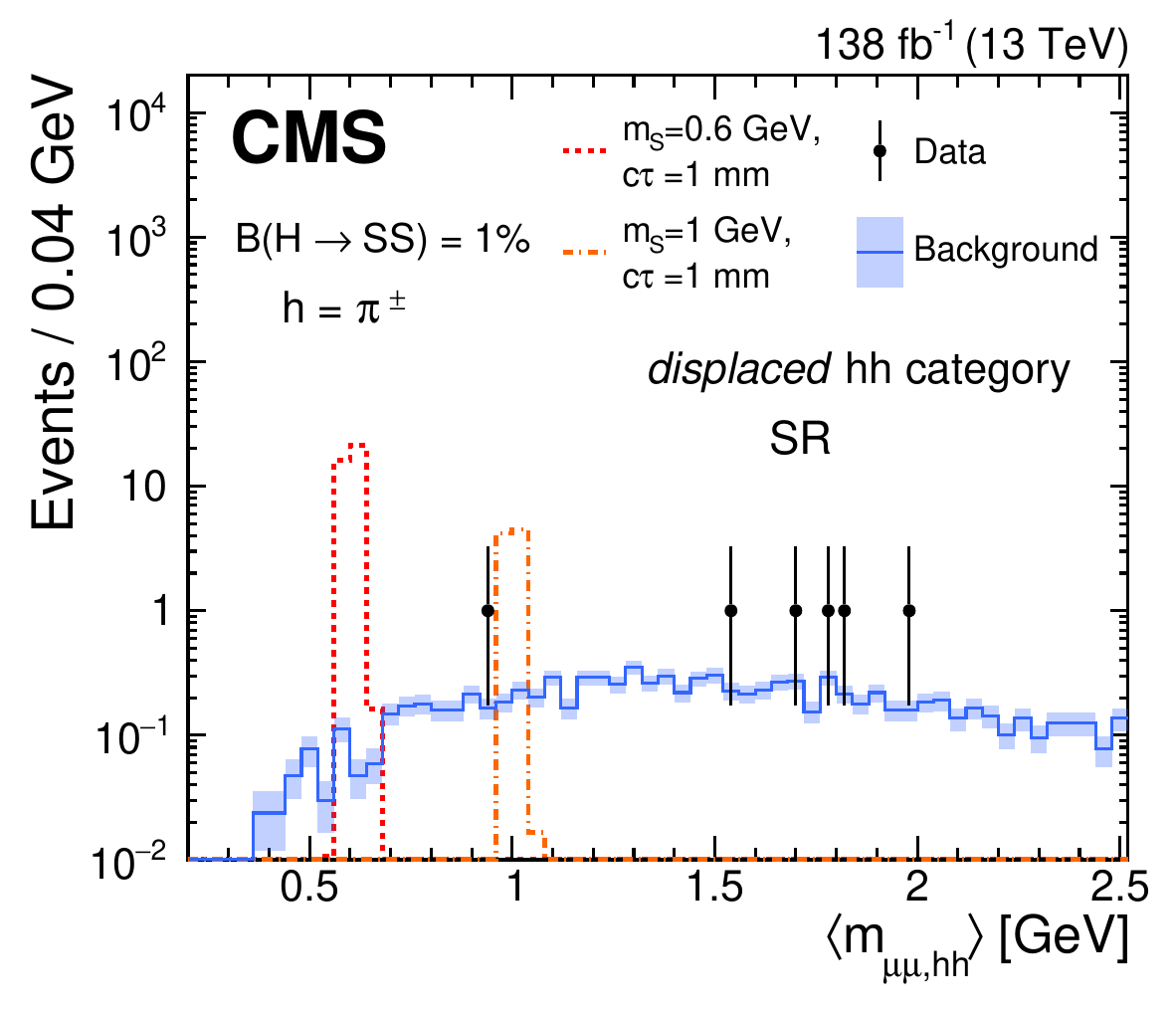}
               \includegraphics[width=0.49\textwidth]{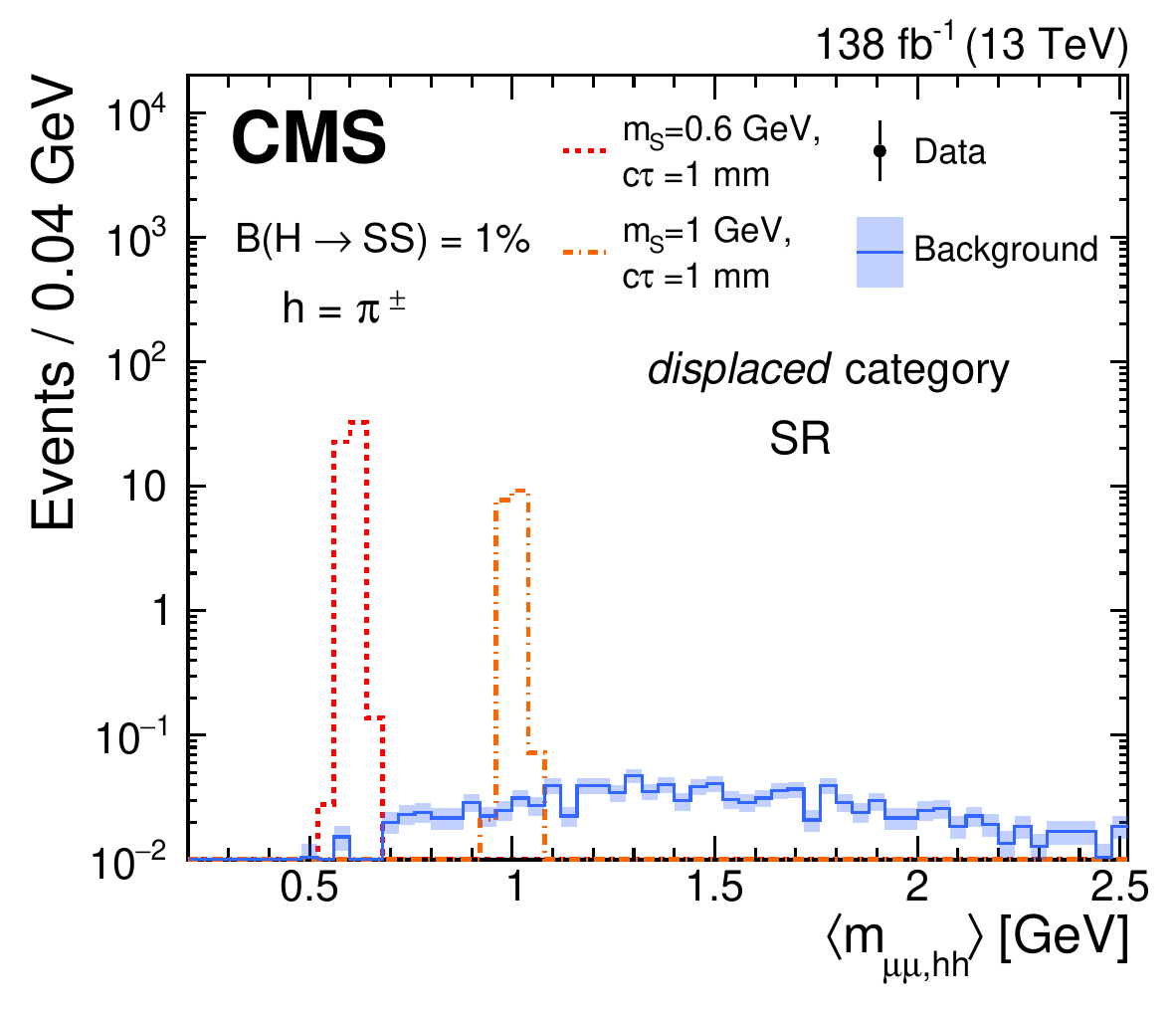}
        \caption{Average diobject invariant mass distributions in the SR for the \textit{prompt} (upper left), \textit{displaced} $\PGm\PGm$ (upper right), \textit{displaced} $\PSh\PSh$ (lower left), and \textit{displaced} (lower right) categories. The event selection criteria of Table~\ref{tab:selection} have been applied. The observed number of events in data are shown as black dots with the uncertainty bars and the background prediction (estimated from the events in the CR as explained in Section~\ref{sec:background}) as blue histograms. Uncertainty bars are not shown for bins with zero entries. Two signal samples are also presented for BSM scalar bosons with $\mS=0.6$ and 1\GeV and $c\tau=1\mm$. The signal yield is scaled to the integrated luminosity, assuming the gluon fusion Higgs boson production cross-section \cite{ggH_cs}, $\mathcal{B}(\PH\to\PS\PS)=1\%$, and $\mathcal{B}(\PS\to\Pgmp\Pgmm)$, and $\mathcal{B}(\PS\to\PGpp\PGpm)$ for $\mS=0.6$ and 1\GeV, according to the values listed in Table~\ref{tab:brfrac}.}
    \label{fig:avgmass_ms0p6_ctau1}
\end{figure}

A summary of the selection criteria is given in Table~\ref{tab:selection}. Figure~\ref{fig:recohiggsmass_ms0p6_ctau1} shows the four-object invariant mass, \mmumuhh, distributions for the data and a reference signal in the four categories separately, after the event selection (as shown in Table~\ref{tab:selection}) is applied. Since the BSM scalar boson targeted in this analysis has a mass $\mS\in[0.4,2]\GeV$, the average diobject invariant mass, denoted as $\mavg=\frac{1}{2}(\mmumu+\mhh)$, is restricted to $[0.2,2.5]\GeV$. For illustration, an exponential fit in the CR is performed to model the background in the SR. Figure~\ref{fig:avgmass_ms0p6_ctau1} shows the $\mavg$ distribution for the observed data, signal and, predicted background events in the SR under the charged pion assumption for the final-state hadrons (i.e., $\Ph=\PGppm$). The details of the background estimation are discussed in the next section. The background is spread over a range of \mavg values while the signal is concentrated in a narrow window. Significant contributions from narrow SM states, such as $\PKzS\to\PGpp\PGpm$ events, were not observed in the data in either the CR or the SR.

This motivates performing the search in sliding windows of \mavg around each considered signal mass hypothesis. A Gaussian function is fit to the signal \mavg distribution to determine the signal mass resolution, $\sigma$. The resolution varies with the considered mass hypotheses and is evaluated for each \mS using samples with the proper decay length $c\tau=1\mm$. Mass windows of $\pm 2.5\sigma$ are then defined around the mass hypotheses to estimate the signal and background yields in the SR for each event category. The expected signal yield in the SR is taken from simulation with corrections applied to account for discrepancies between data and simulation. This includes corrections for the pileup reweighting, trigger efficiency, muon reconstruction efficiency, and secondary vertex reconstruction efficiency.

\section{Background estimation} \label{sec:background}

The background contribution in the SR is estimated from data in the CR. An exponential fit to the CR provides a reliable model for the background in the SR. The transfer factor between the sidebands and the SR is defined as $f_{\text{SR}/\text{CR}}=N^{\text{SR}}_{\text{fit}}/N^{\text{CR}}_{\text{fit}}$, where $N^{\text{SR}}_{\text{fit}}$ ($N^{\text{CR}}_{\text{fit}}$) is the number of events in the SR (CR) extracted from the fit. The events from the CR are then reweighted with the transfer factor to estimate the background contribution in the SR. 

The above procedure is applied to the \textit{prompt} category, as a large number of events are present in the CR, even after imposing the \mavg window constraint corresponding to each signal hypothesis. The resulting background estimation in the SR for the \textit{prompt} category is shown as the blue histogram in Fig.~\ref{fig:avgmass_ms0p6_ctau1} (upper left).

For the non-\textit{prompt} categories, the CR contains very few events within each signal mass window, resulting in an unreliable estimate of the background. In particular, there are nearly zero events in the CR for the displaced category even before the \mavg window is applied, as shown in Fig.~\ref{fig:recohiggsmass_ms0p6_ctau1} (lower right). Therefore, a looser CR is defined for the non-\textit{prompt} categories by relaxing some selection criteria.

This proceeds over several steps. First, non-\textit{prompt} categories are merged into a single category, since the background shape was found to be similar across those three different categories. At this stage, the factor associated with the fraction of events of any particular category that reside in the new merged non-\textit{prompt} category is calculated as $f_{1}=\frac{N_{\text{category}}}{N_{\text{merged category}}}$. Next, the isolation requirements are removed. To estimate the transfer factor between the regions with and without the isolation requirements, the two corresponding \mavg distributions are compared for the background events lying in the sidebands of the \mmumuhh distribution. A fit using a Landau function is performed on the \mavg distribution for the case with no isolation requirement applied. The same function with free normalization is fitted to the event distribution where the isolation criteria are applied to obtain the transfer factor $f_{2}={N_{\text{isolation, merged}}}/{N_{\text{no isolation, merged}}}$. It was verified that the functional form of the \mavg distribution is not affected by the isolation requirement and is consistent across the non-$prompt$ categories. Independent Landau fits performed before and after relaxing the selection criteria for loose CR were found to be compatible. The transfer factor $f_{\text{SR}/\text{CR}}$ for this loose CR is then estimated with an exponential fit. The loose CR events, weighted with $f_1$, $f_2$, and $f_{\text{SR}/\text{CR}}$ yield the background prediction shown as the blue histograms in Fig.~\ref{fig:avgmass_ms0p6_ctau1} (top right, bottom left, bottom right) for each of the non-\textit{prompt} categories. We observe that data are in agreement with the background estimation.

\begin{figure}[tbh!]
    \centering
            \includegraphics[width=0.49\textwidth]{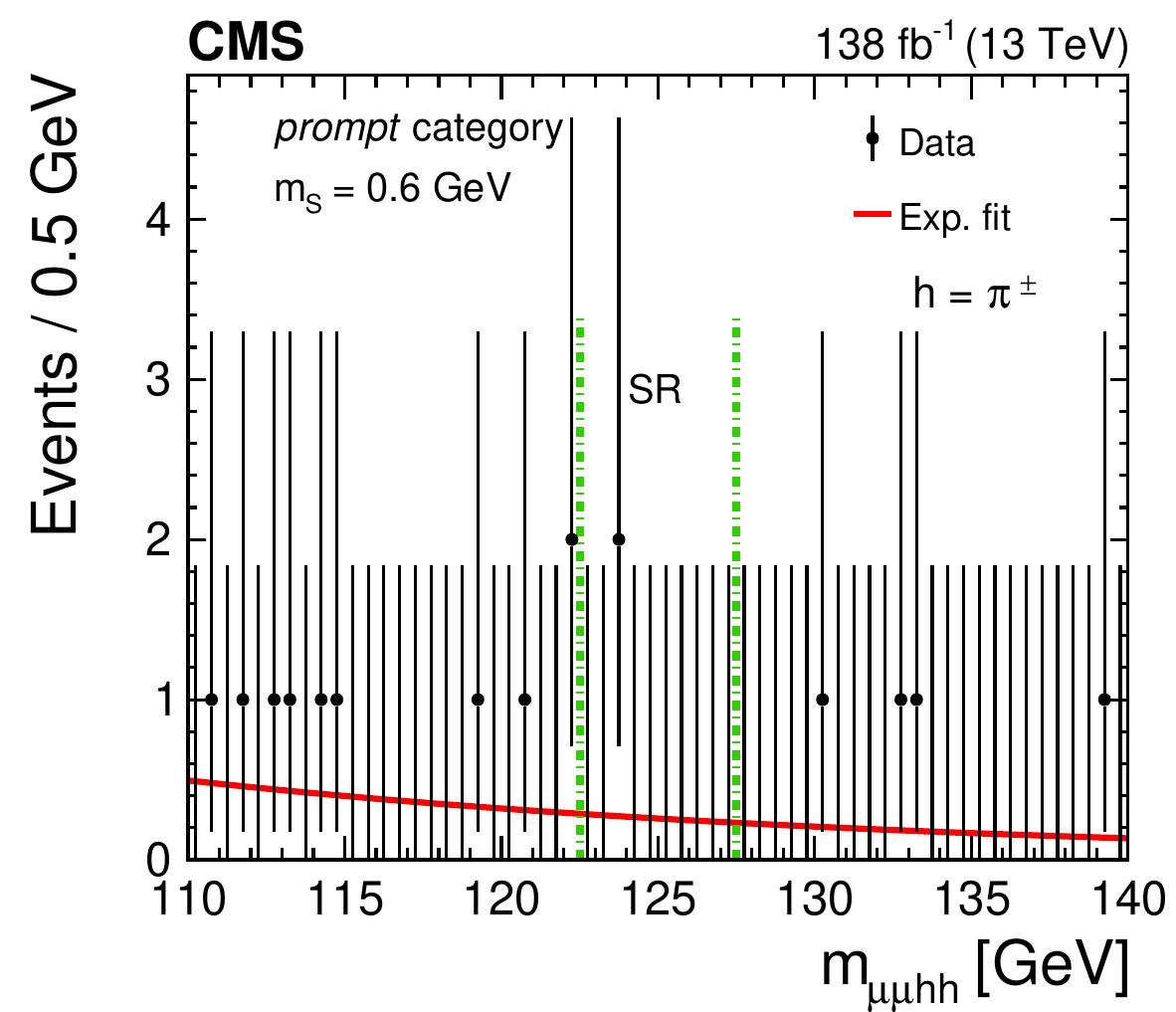}
           \includegraphics[width=0.49\textwidth]{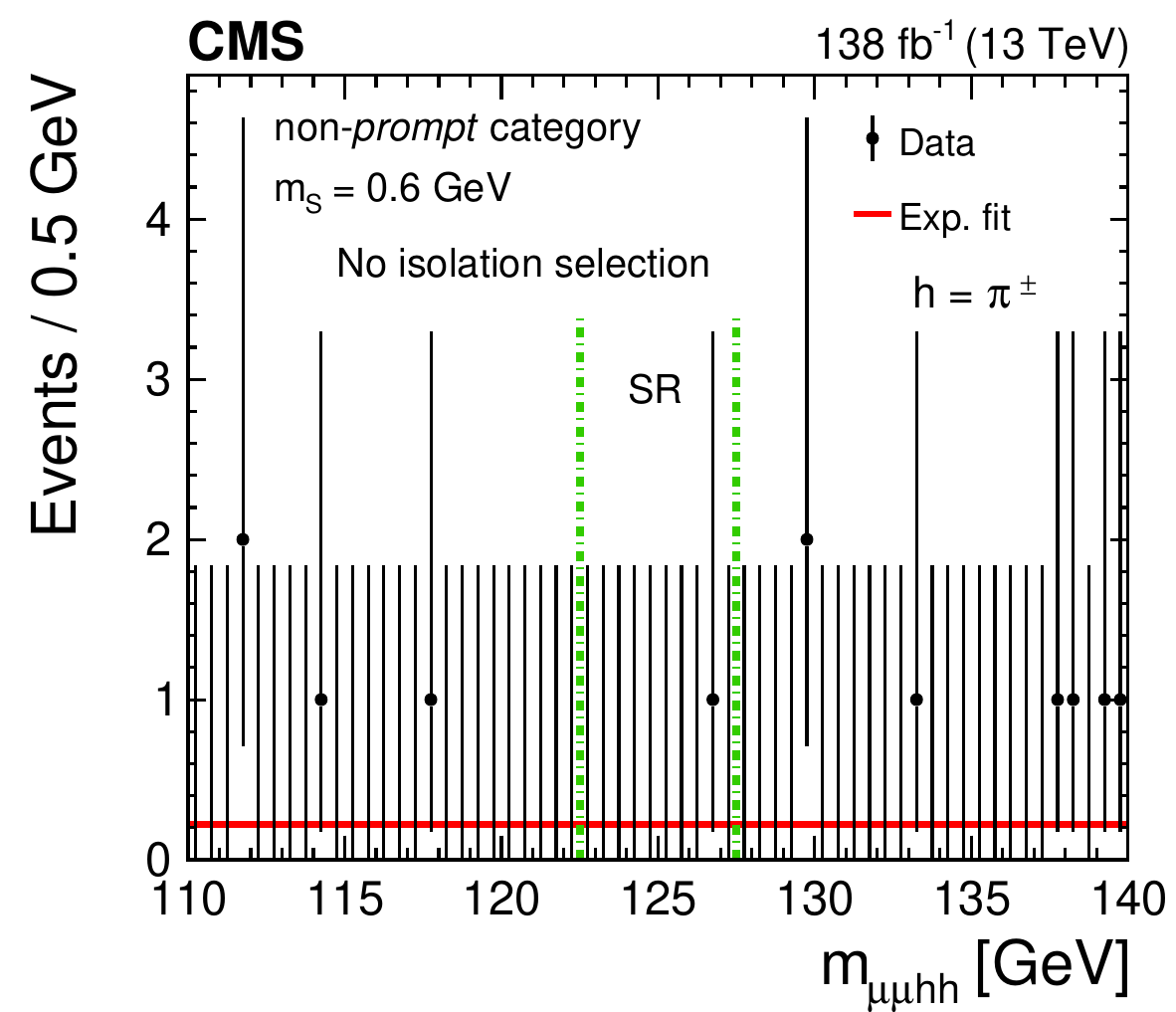}
	\caption{The four-object invariant mass distribution for the 2016--2018 dataset in the \textit{prompt} (\cmsLeft) and merged non-\textit{prompt} (\cmsRight) categories after the application of the event selection listed in Table~\ref{tab:selection} and the mass window for $\mS=0.6\GeV$. The dots with uncertainty bars represent the data. The red curve is the result of the exponential fit to the CR. The green dash-dotted vertical lines delineate the SR. The distributions for the merged non-\textit{prompt} nonisolated category are not weighted by the relevant transfer factors.}
    \label{fig:expofit_dispmerged}
\end{figure}

Finally, the expected background and signal yields are extracted after applying the \mavg window corresponding to each signal mass hypothesis. As an example, Fig.~\ref{fig:expofit_dispmerged} shows the \mmumuhh distribution in the \textit{prompt} category (\cmsLeft) and merged non-\textit{prompt} category (\cmsRight), for the \mavg window at $\mS=0.6\GeV$. The merged non-\textit{prompt} category shows the unweighted distribution without transfer factors for events passing the relaxed criteria discussed before. Table~\ref{tab:bkgyield} shows the signal yield for $\mS=0.6\GeV$ and $c\tau=1\mm$ (upper), and $\mS=1.6\GeV$ and $c\tau=1\mm$ (lower), the background prediction, and the observed data.

The background estimation is performed individually for each era of data taking to account for the variations in the number of background events arising from differences in trigger requirements, such as muon $\pt$ thresholds, and pileup conditions between eras. The statistical uncertainty in the background estimation is within 20--100 (15--70)\% for the \textit{prompt} (merged non-\textit{prompt}) category in each era. The uncertainty in the transfer factors for the non-\textit{prompt} categories is included, as described in the next section. 

\begin{table}[tbh!]
          \centering
	    \topcaption{Number of observed events, background predictions for the 2016--2018 dataset in the SR under the BSM scalar mass hypothesis, $\mS=0.6\GeV$ (upper) and $\mS=1.6\GeV$ (lower), and the expected signal yield assuming $\mathcal{B}(\PH\to \PS\PS)=1\%$.}
	    \label{tab:bkgyield}
        \begin{tabular}{lccc}
            Region & Data & Background prediction & $\mS=0.6\GeV$, $c\tau=1\mm$ \\ \hline
            \textit{Prompt} & 2 & $2.0\pm 0.5$ & $42\pm 6$\\ 
            \textit{Displaced} $\PGm\PGm$  & 0 & $0.15\pm 0.04$ & $64\pm 13$\\ 
            \textit{Displaced} $\Ph\Ph$  & 0 & $0.08\pm 0.02$ & $34\pm 6$\\ 
            \textit{Displaced} & 0 & $0.010\pm 0.003$ & $51\pm 11$ \\
        \end{tabular} 
      
    \vspace{1em}

        \begin{tabular}{lccc}
            Region & Data &  Background prediction & $\mS=1.6\GeV$, $c\tau=1\mm$\\ \hline
            \textit{Prompt} & 1 & $4.7\pm 0.9$ & $94\pm 14$ \\ 
            \textit{Displaced} $\PGm\PGm$ & 0 & $1.4 \pm 0.1$ & $203\pm 41$\\ 
            \textit{Displaced} $\Ph\Ph$ & 0 & $0.52\pm 0.05$ & $106\pm 19$ \\ 
            \textit{Displaced} & 0 & $0.15\pm 0.02$ & $190\pm 40$\\
        \end{tabular}
\end{table}

\section{Systematic uncertainties} \label{sec:systematics}

The background prediction uncertainty is dominated by the statistical uncertainty from the limited number of events in the CR. Alternate functional forms to the exponential model, including a power law and linear functions, were explored. The variation in the resulting background predictions, propagated via the transfer factor $f_{\text{SR}/\text{CR}}$, was found to be very small compared to the statistical uncertainty, and was neglected. In the merged non-$prompt$ category, additional sources of uncertainties are included to account for the multiple transfer factors. The uncertainty in $f_{1}$ is up to 100\% in the \textit{displaced} category and arises from the nearly zero events in the sideband after applying the event selection. The transfer factor $f_{2}$ has an associated uncertainty in a range of 10--30\%. 

Several experimental and theoretical uncertainties affect the signal prediction. The leading theoretical uncertainty, which dominates the total signal uncertainty, comes from the change in the Higgs boson \pt distribution arising from renormalization and factorization scale variations. This uncertainty is estimated by varying the renormalization and factorization scales up and down by a factor of two, excluding the two extreme variations, and amounts to $\approx$17\%.

The leading experimental uncertainties come from the corrections applied to the secondary vertex reconstruction of displaced vertices and the isolation requirements applied on the hadrons. Corrections to the displaced-vertex reconstruction efficiency are derived as a function of displacement and momentum by using decays of neutral kaons to pairs of charged pions in data, based on the methodology of Ref.~\cite{CMS-DP-2024-010}. The uncertainty ranges between 5--40\%, with the larger one present for the non-\textit{prompt} categories. A similar methodology is adopted to derive the uncertainties in the signal mass window constructed for the event selection. The corrections and uncertainties for the displaced-vertex reconstruction efficiency are applied to the dihadron vertices. The uncertainties for the signal mass window are applied to both the dimuon and dihadron vertices. A conservative estimate of the uncertainty for the muon and hadron isolation requirements is applied. This uncertainty is calculated separately for the muon and hadron isolation thresholds. In the \textit{prompt} category, the uncertainty associated with the muon (hadron) isolation is 5--12 (6--16)\%. In the non-\textit{prompt} categories, the isolation requirements are much looser, leading to a highly efficient selection, and the uncertainty in the muon (hadron) isolation leads to a 2.5 (10)\% uncertainty in the signal yield. The uncertainty in the reconstruction of hadron tracks is $\sim$5\%~\cite{CMS-DP-2022-012}. Uncertainties in the muon identification and reconstruction of close-by muons in the tracker amounts to less than 1.2\% per muon~\cite{HIG-21-004}. Additionally, a 5\% uncertainty per muon is assigned to account for the reconstruction of displaced muons~\cite{CMS:2022qej}.

Corrections to the trigger efficiency in simulation are derived with a "tag-and-probe" method~\cite{CMS:2010svw} using the \JPsi resonance, which closely emulate the signal features. The related trigger uncertainty is found to be 5--13\%. The pileup modeling uncertainties amount to 1\%. The integrated luminosities for the 2016, 2017, and 2018 data-taking years have 1--3\% respective uncertainties~\cite{CMS-LUM-17-003,CMS-PAS-LUM-17-004,CMS-PAS-LUM-18-002}, with an overall uncertainty of 1.6\%. Finally, the uncertainty due to the size of the signal samples is also included and varies from 2 to 15\%, with larger values towards the larger $c\tau$ hypotheses. The uncertainties are summarized in Table \ref{tab:unc_summary}.

\begin{table}[thb!]
    \centering
    \topcaption{Summary of systematic uncertainties.}\label{tab:unc_summary}
    \cmsTable{
    \begin{tabular}{llll}
         Systematic uncertainty & Value & Applied to & Comments\\ \hline
         \multirow{2}{*}{Background event count}  & \multirow{2}{*}{15--100\%} & \multirow{2}{*}{Background} &event-level, based on scalar boson mass\\
         & & & \& category \\
         Transfer factor $f_1$  & 15--100\% & Background & only non-\textit{prompt} categories\\
         Transfer factor $f_2$  & 10--30\% & Background & only non-\textit{prompt} categories\\ 
         Higgs boson \pt spectrum & 17\% & Signal & event-level \\
         SV reco. & 5--40\% & Signal & dihadron SV, based on category\\
         Reco. scalar boson mass reso. & 1--10\% & Signal & per scalar boson mass, based on category\\
         Muon isolation & 2.5--12\% & Signal & event-level, based on scalar boson mass\\
         Hadron isolation & 6--16\% & Signal & event-level, based on scalar boson mass \\
         Reco. hadronic tracks & 4--5\% & Signal & per hadron\\
         Muon identification & 0.6\% & Signal & per muon\\
         Reco. close muons & 1\% & Signal & event-level\\
         Reco. displaced muons & 5\% & Signal & per muon\\
         Trigger & 5--13\% & Signal & event-level\\
         Pileup modeling & 1\% & Signal & event-level\\
         Integrated luminosity & 1--3\% & Signal & event-level\\
         MC event count & 2--15\% & Signal & event-level \\
            \end{tabular}
    }
\end{table}

\section{Results} \label{sec:results}

With the application of the selection and the related uncertainties, the predicted background count is used to perform a statistical interpretation of the results. Under the background-only hypothesis, 95\% confidence level (\CL) upper limits on the branching fraction of $\mathcal{B}(\PH\to\PS\PS)$ are calculated using the \CLs method \cite{Junk_1999,2015_CLs} for the benchmark model. The contributions from Higgs boson via vector boson fusion and production in association with a vector boson are evaluated and accounted for as well, and increase the signal yields (across different \mS and decay lengths) by about 9.1\%. The limits are derived using a modified frequentist approach, with the likelihood ratio as the test statistic. Given the small event count, pseudo-experiments are used to obtain the distribution of the test statistic. The results are determined with the CMS statistical analysis tool \textsc{Combine}~\cite{CMS:2024onh} which is based on the \textsc{RooFit}~\cite{Verkerke:2003ir} and \textsc{RooStats}~\cite{Moneta:2010pm} frameworks. 

\begin{figure}[tbh!] 
                \centering
                \includegraphics[width=0.49\textwidth]{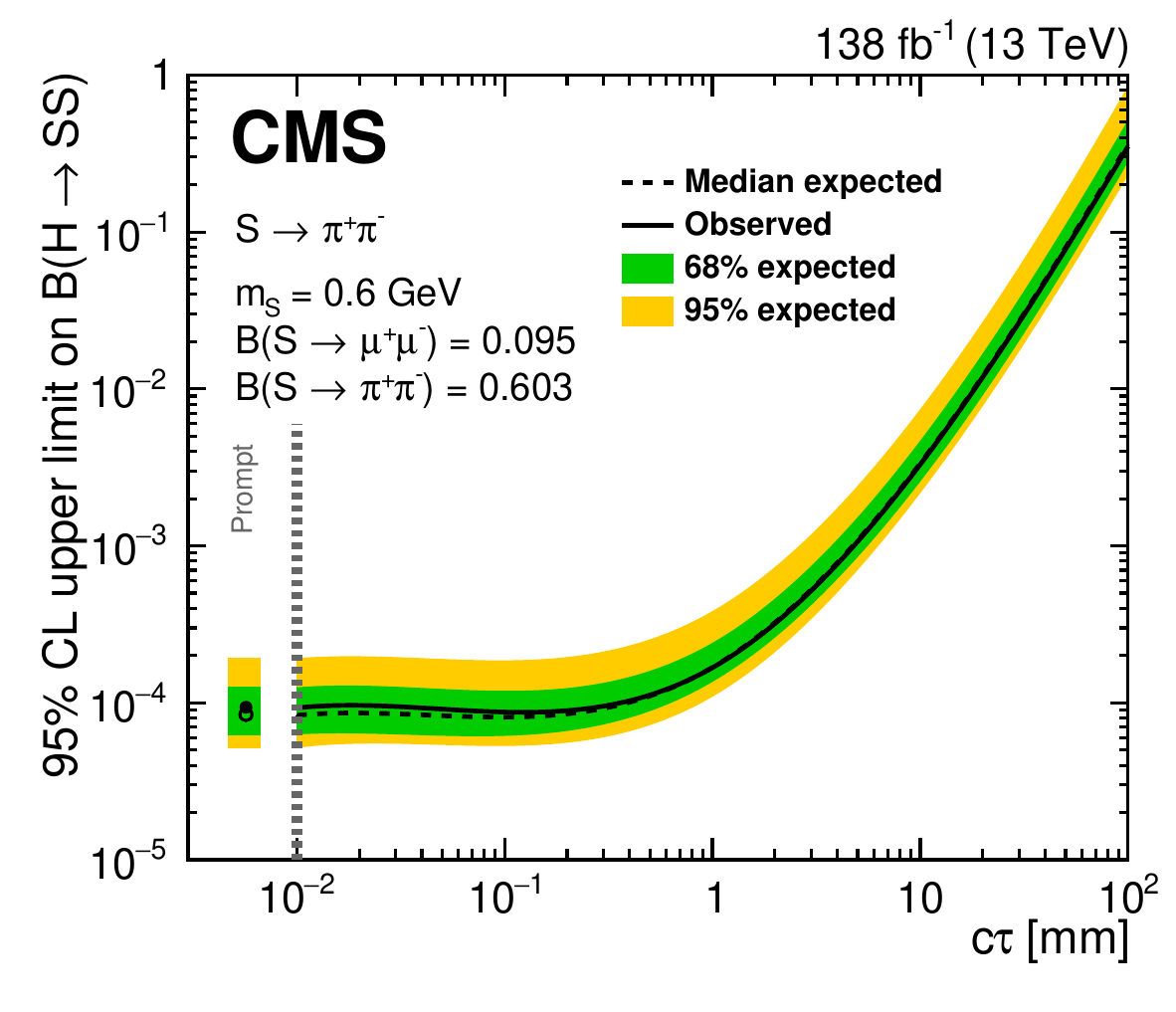}
                    \includegraphics[width=0.49\textwidth]{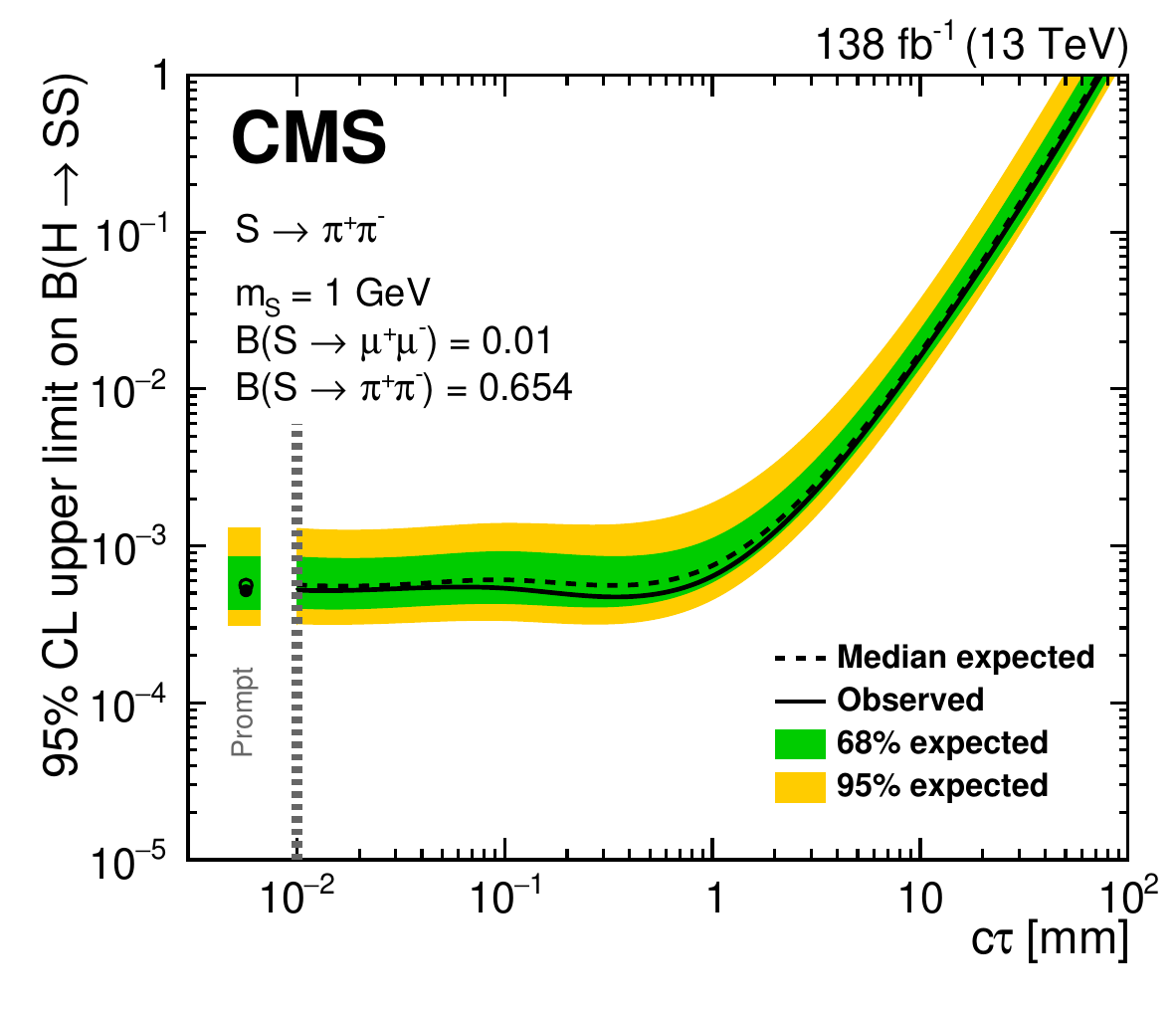}
        \caption{Observed (solid black line) and expected (dashed black line) exclusion limits on the branching fraction $\mathcal{B}(\PH\to\PS\PS)$ as functions of signal proper decay length $c\tau$ for $\mS=0.6\GeV$ (\cmsLeft) and 1\GeV (\cmsRight) for the minimal extension of the SM Higgs sector \cite{Winkler:2018qyg}, assuming $\mathcal{B}(\PS\to\PGmp\PGmm)$ and $\mathcal{B}(\PS\to\PGpp\PGpm)$, as in Table~\ref{tab:brfrac}. The inner green (outer yellow) indicates the region containing 68 (95)\% of the limits. The limits are obtained using the combination of all data-taking eras and the four different displacement categories.}
        \label{fig:lims_ms_pion}
\end{figure}

\begin{figure}[tbh!]
                \centering
            \includegraphics[width=0.49\textwidth]{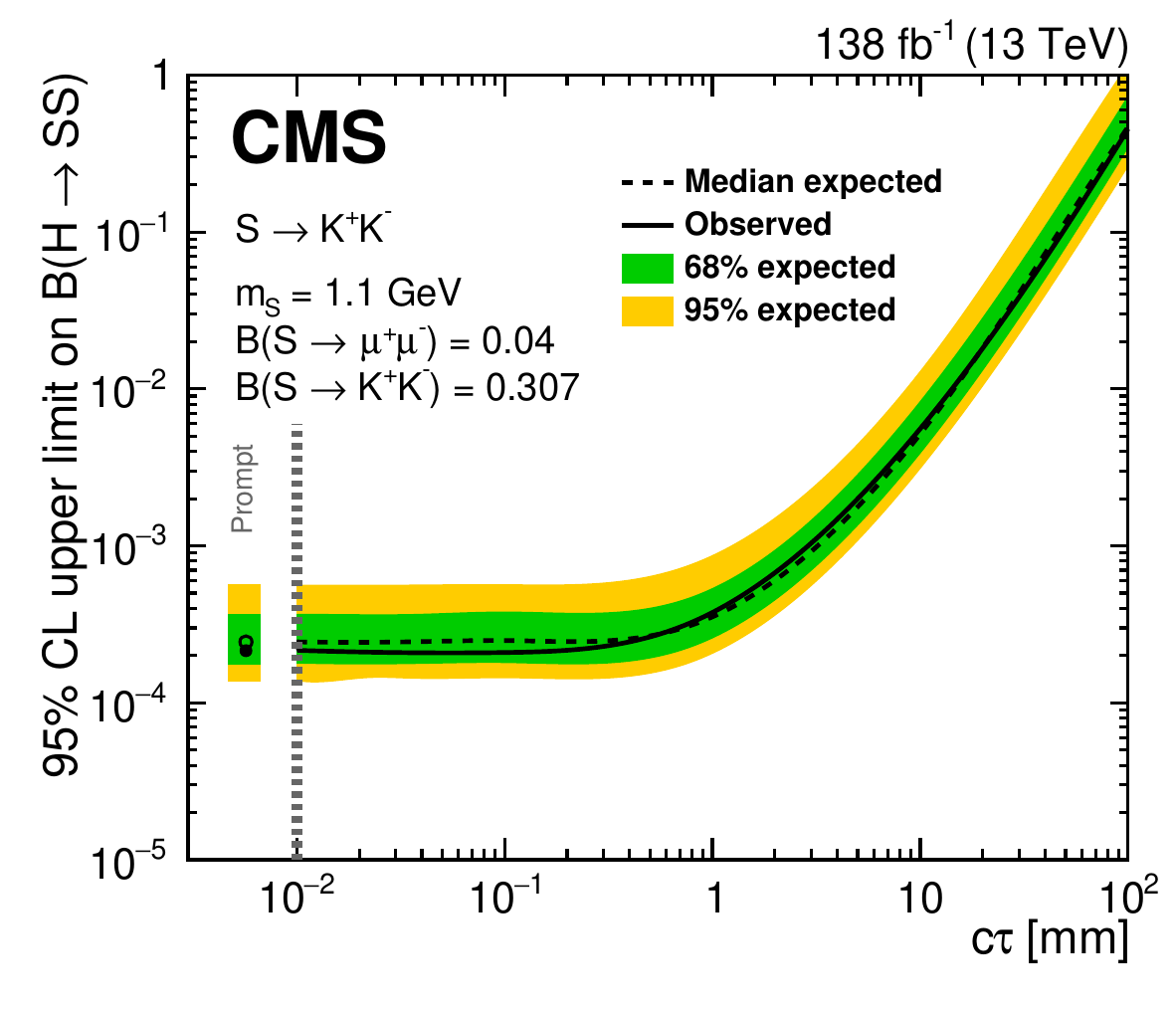}
            \includegraphics[width=0.49\textwidth]{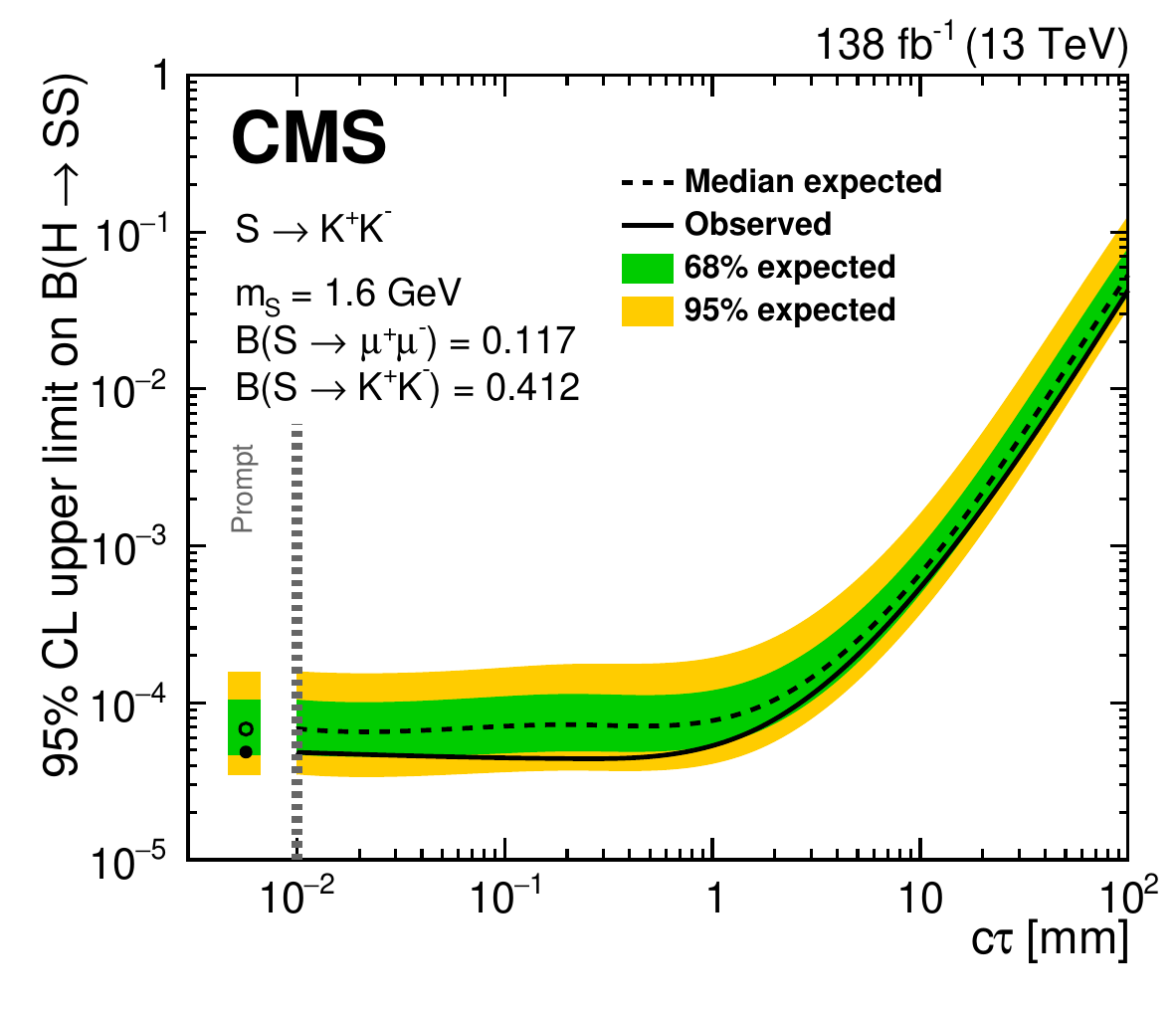}
    \caption{Observed (solid black line) and expected (dashed black line) exclusion limits on the branching fraction $\mathcal{B}(\PH\to\PS\PS)$ as functions of signal proper decay length $c\tau$ for $\mS=1.1\GeV$ (\cmsLeft) and $1.6\GeV$ (\cmsRight) for the minimal extension of the SM Higgs sector~\cite{Winkler:2018qyg}, assuming $\mathcal{B}(\PS\to\PGmp\PGmm)$ and $\mathcal{B}(\PS\to \PKp\PKm)$, as in Table~\ref{tab:brfrac}. The inner green (outer yellow) indicates the region containing 68 (95)\% of the limits. The limits are obtained using the combination of all data-taking eras and the four different displacement categories.}
    \label{fig:lims_ms_kaon}
\end{figure}

\begin{figure}[tbhp!]
        \centering
        \includegraphics[width=0.8\textwidth]{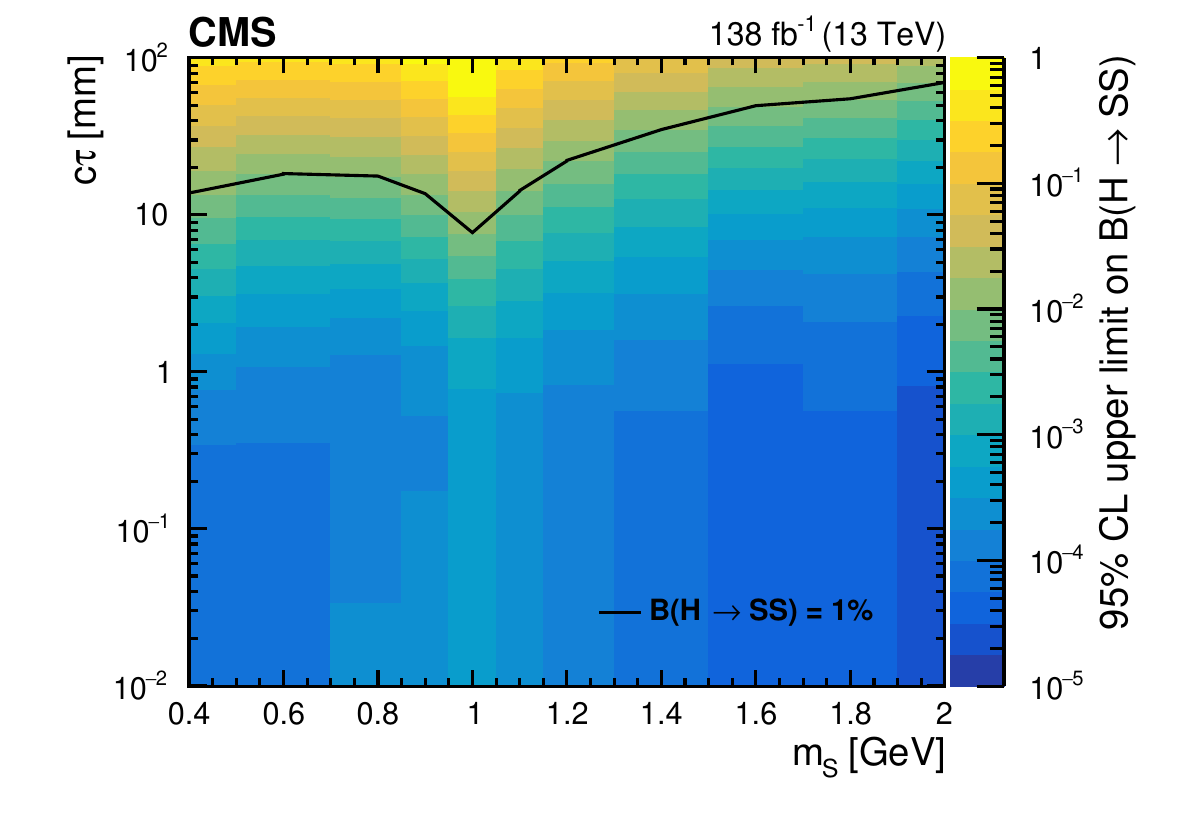}
    \caption{Observed upper limit on the branching fraction $\mathcal{B}(\PH\to\PS\PS)$ as a function of signal mass and proper decay length $c\tau$ for the minimal extension of the SM Higgs sector \cite{Winkler:2018qyg}, assuming $\mathcal{B}(\PS\to\PGmp\PGmm)$, $\mathcal{B}(\PS\to \PKp\PKm)$, and $\mathcal{B}(\PS\to\PGpp\PGpm)$, as in Table~\ref{tab:brfrac}. The area below the solid black line denotes the region where the limits are smaller than 1\%.}
    \label{fig:lims_2d}
\end{figure}

Figures~\ref{fig:lims_ms_pion} and \ref{fig:lims_ms_kaon} show the observed and expected limits as functions of the decay length for different \mS hypotheses. In Fig.~\ref{fig:lims_ms_pion}, the limits for $\mS=0.6$ and 1\GeV are presented, using the final state of two muons and two charged pions. Below $c\tau\lesssim 0.1\mm$, the limits start to plateau with respect to the decay length. Most of the events in this region fall within the \textit{prompt} category, which benefits from efficient reconstruction and triggering capabilities. This also indicates that below a certain decay length, the events approach the prompt-like signature and the difference in the reconstructed vertex displacement between the decay lengths is negligible. Beyond $c\tau=0.1\mm$, the limits start to worsen as the vertices are more displaced, and the events are harder to trigger on and reconstruct. Similar behavior is observed in Fig.~\ref{fig:lims_ms_kaon}, where the limits for $\mS=1.1$ and 1.6\GeV are presented, using the final state of two muons and two charged kaons. The observed limits over the full mass range of 0.4--2\GeV are shown in Fig.~\ref{fig:lims_2d} as functions of \mS and $c\tau$. Improved sensitivity is achieved at higher masses because of lower backgrounds, while a drop near 1\GeV is observed resulting from a smaller $\mathcal{B}(\PS\to\Pgm\Pgm)$. For all other mass hypotheses considered, a limit on $\mathcal{B}(\PH\to\PS\PS)$ of the order of $10^{-3}$ ($10^{-4}$) for proper decay lengths up to $c\tau\sim10\mm$ (1\mm) is achieved.

\clearpage
\section{Summary} \label{sec:summary}
An exclusive search for light exotic scalar bosons has been presented using proton-proton collision data collected in 2016--2018 at $\sqrt{s}=13\TeV$, and corresponding to an integrated luminosity of 138\fbinv. The analysis accesses a unique hadronic decay mode where the low mass of the scalar boson restricts hadronization and dominantly allows decays to only a pair of light hadrons. The Higgs boson-mediated production of a pair of light scalar bosons decaying within the CMS trackers is sought in the final state of two muons and two charged hadrons. Scalar boson masses below 2\GeV with a proper decay length $c\tau$ up to 100\mm are probed. The boosted topology, the isolation, and mass constraints between the reconstructed dimuon and dihadron resonances are used to reduce the background. Considering the benchmark signal model~\cite{Gershtein:2020mwi}, this search excludes at 95\% confidence level branching fractions of the Higgs boson to light scalar bosons greater than $\mathcal{O}(10^{-4})$ for $c\tau$ up to $\sim 1\mm$ and covers a largely unexplored parameter space of light long-lived particles.

\begin{acknowledgments}
We congratulate our colleagues in the CERN accelerator departments for the excellent performance of the LHC and thank the technical and administrative staffs at CERN and at other CMS institutes for their contributions to the success of the CMS effort. In addition, we gratefully acknowledge the computing centers and personnel of the Worldwide LHC Computing Grid and other centers for delivering so effectively the computing infrastructure essential to our analyses. Finally, we acknowledge the enduring support for the construction and operation of the LHC, the CMS detector, and the supporting computing infrastructure provided by the following funding agencies: SC (Armenia), BMBWF and FWF (Austria); FNRS and FWO (Belgium); CNPq, CAPES, FAPERJ, FAPERGS, and FAPESP (Brazil); MES and BNSF (Bulgaria); CERN; CAS, MoST, and NSFC (China); MINCIENCIAS (Colombia); MSES and CSF (Croatia); RIF (Cyprus); SENESCYT (Ecuador); ERC PRG and PSG, TARISTU24-TK10 and MoER TK202 (Estonia); Academy of Finland, MEC, and HIP (Finland); CEA and CNRS/IN2P3 (France); SRNSF (Georgia); BMFTR, DFG, and HGF (Germany); GSRI (Greece); MATE and NKFIH (Hungary); DAE and DST (India); IPM (Iran); SFI (Ireland); INFN (Italy); MSIT and NRF (Republic of Korea); MES (Latvia); LMTLT (Lithuania); MOE and UM (Malaysia); BUAP, CINVESTAV, CONACYT, LNS, SEP, and UASLP-FAI (Mexico); MOS (Montenegro); MBIE (New Zealand); PAEC (Pakistan); MSHE, NSC, and NAWA (Poland); FCT (Portugal);  MESTD (Serbia); MICIU/AEI and PCTI (Spain); MOSTR (Sri Lanka); Swiss Funding Agencies (Switzerland); MST (Taipei); MHESI (Thailand); TUBITAK and TENMAK (T\"{u}rkiye); NASU (Ukraine); STFC (United Kingdom); DOE and NSF (USA).

\begin{sloppypar}
\setlength\emergencystretch{\hsize}
\hyphenation{Rachada-pisek} Individuals have received support from the Marie-Curie program and the European Research Council and Horizon 2020 Grant, contract Nos.\ 675440, 724704, 752730, 758316, 765710, 824093, 101115353, 101002207, 101001205, and COST Action CA16108 (European Union); the Leventis Foundation; the Alfred P.\ Sloan Foundation; the Alexander von Humboldt Foundation; the Science Committee, project no. 22rl-037 (Armenia); the Fonds pour la Formation \`a la Recherche dans l'Industrie et dans l'Agriculture (FRIA) and Fonds voor Wetenschappelijk Onderzoek contract No. 1228724N (Belgium); the Beijing Municipal Science \& Technology Commission, No. Z191100007219010, the Fundamental Research Funds for the Central Universities, the Ministry of Science and Technology of China under Grant No. 2023YFA1605804, the Natural Science Foundation of China under Grant No. 12535004, and USTC Research Funds of the Double First-Class Initiative No.\ YD2030002017 (China); the Ministry of Education, Youth and Sports (MEYS) of the Czech Republic; the Shota Rustaveli National Science Foundation (Georgia); the Deutsche Forschungsgemeinschaft (DFG), among others, under Germany's Excellence Strategy -- EXC 2121 ``Quantum Universe" -- 390833306, and under project number 400140256 - GRK2497; the Hellenic Foundation for Research and Innovation (HFRI), Project Number 2288 (Greece); the Hungarian Academy of Sciences, the New National Excellence Program - \'UNKP, the NKFIH research grants K 131991, K 138136, K 143460, K 143477, K 147557, K 146913, K 146914, K 147048, TKP2021-NKTA-64, and 2025-1.1.5-NEMZ\_KI-2025-00004, and MATE KKP and KKPCs Research Excellence and Flagship Research Groups grants (Hungary); the Council of Science and Industrial Research, India; ICSC -- National Research Center for High Performance Computing, Big Data and Quantum Computing, FAIR -- Future Artificial Intelligence Research, and CUP I53D23001070006 (Mission 4 Component 1), funded by the NextGenerationEU program, the Italian Ministry of University and Research (MUR) under Bando PRIN 2022 -- CUP I53C24002390006, PRIN PRIMULA 2022RBYK7T (Italy); the Latvian Council of Science; the Ministry of Science and Higher Education, project no. 2022/WK/14, and the National Science Center, contracts Opus 2021/41/B/ST2/01369, 2021/43/B/ST2/01552, 2023/49/B/ST2/03273, and the NAWA contract BPN/PPO/2021/1/00011 (Poland); the Funda\c{c}\~ao para a Ci\^encia e a Tecnologia (Portugal); the National Priorities Research Program by Qatar National Research Fund;  MICIU/AEI/10.13039/501100011033, ERDF/EU, ``European Union NextGenerationEU/PRTR", projects PID2022-142604OB-C21, PID2022-139519OB-C21, PID2023-147706NB-I00, PID2023-148896NB-I00, PID2023-146983NB-I00, PID2023-147115NB-I00, PID2023-148418NB-C41, PID2023-148418NB-C42, PID2023-148418NB-C43, PID2023-148418NB-C44, PID2024-158190NB-C22, RYC2021-033305-I, RYC2024-048719-I, CNS2023-144781, CNS2024-154769 and Plan de Ciencia, Tecnolog{\'i}a e Innovaci{\'o}n de Asturias, Spain; the Chulalongkorn Academic into Its 2nd Century Project Advancement Project, the National Science, Research and Innovation Fund program IND\_FF\_68\_369\_2300\_097, and the Program Management Unit for Human Resources \& Institutional Development, Research and Innovation, grant B39G680009 (Thailand); the Eric \& Wendy Schmidt Fund for Strategic Innovation through the CERN Next Generation Triggers project under grant agreement number SIF-2023-004; the Kavli Foundation; the Nvidia Corporation; the SuperMicro Corporation; the Welch Foundation, contract C-1845; and the Weston Havens Foundation (USA).
\end{sloppypar}
\end{acknowledgments}

\bibliography{auto_generated} 
\cleardoublepage \appendix\section{The CMS Collaboration \label{app:collab}}\begin{sloppypar}\hyphenpenalty=5000\widowpenalty=500\clubpenalty=5000\cmsinstitute{Yerevan Physics Institute, Yerevan, Armenia}
{\tolerance=6000
A.~Hayrapetyan, V.~Makarenko\cmsorcid{0000-0002-8406-8605}, A.~Tumasyan\cmsAuthorMark{1}\cmsorcid{0009-0000-0684-6742}
\par}
\cmsinstitute{Institut f\"{u}r Hochenergiephysik, Vienna, Austria}
{\tolerance=6000
W.~Adam\cmsorcid{0000-0001-9099-4341}, L.~Benato\cmsorcid{0000-0001-5135-7489}, T.~Bergauer\cmsorcid{0000-0002-5786-0293}, M.~Dragicevic\cmsorcid{0000-0003-1967-6783}, P.S.~Hussain\cmsorcid{0000-0002-4825-5278}, M.~Jeitler\cmsAuthorMark{2}\cmsorcid{0000-0002-5141-9560}, N.~Krammer\cmsorcid{0000-0002-0548-0985}, A.~Li\cmsorcid{0000-0002-4547-116X}, D.~Liko\cmsorcid{0000-0002-3380-473X}, M.~Matthewman, J.~Schieck\cmsAuthorMark{2}\cmsorcid{0000-0002-1058-8093}, R.~Sch\"{o}fbeck\cmsAuthorMark{2}\cmsorcid{0000-0002-2332-8784}, M.~Shooshtari\cmsorcid{0009-0004-8882-4887}, M.~Sonawane\cmsorcid{0000-0003-0510-7010}, W.~Waltenberger\cmsorcid{0000-0002-6215-7228}, C.E.~Wulz\cmsAuthorMark{2}\cmsorcid{0000-0001-9226-5812}
% ...
\par}
$^{1}$Also at Yerevan State University, Yerevan, Armenia\\
$^{2}$Also at Technische Universit\"{a}t Wien, Vienna, Austria\\
% ...
\end{sloppypar}
\end{document}